\documentclass[]{aa}  

\usepackage{graphicx}
\usepackage{xcolor}
\usepackage{tikz}
\usetikzlibrary{positioning,arrows.meta,shapes.geometric,shapes.symbols}
\usepackage[colorlinks=true, citecolor=blue]{hyperref}

\usepackage{physics}
\usepackage{txfonts}

\newcommand{\bandpass}{\mathbb{B}_{\nu}}
\newcommand{\slope}{\mathbb{S}_{\nu}}
\newcommand{\slopeeps}{\mathbb{S}_{\nu}^{\epsilon}}
\newcommand{\amtn}{-\tau_\nu \, AM}
\newcommand{\tatm}{T^{atm}_\nu}
\newcommand{\tref}{T^{ref}_\nu}
\newcommand{\tstr}{T^{str}_\nu}
\newcommand{\ttot}{T^{tot}_\nu}
\newcommand{\tcos}{T^{cos}_{i,\nu}}

\begin{document} 

\title{CONCERTO: forward modelling of interferograms for calibration}

\author{A. Lundgren\inst{\ref{lam}} 
\and A. Beelen\inst{\ref{lam}}
\and G. Lagache\inst{\ref{lam}}
\and F.-X. D$\rm\acute{e}$sert \inst{\ref{ipag}}
\and A. Fasano\inst{\ref{iac}, \ref{ull}}
\and J. Macias-Perez\inst{\ref{lpsc}}
\and A. Monfardini\inst{\ref{neel}}
\and P. Ade\inst{\ref{aig}}
\and M. Aravena\inst{\ref{Santiago}}
\and E. Barria\inst{\ref{neel}}
\and A. Benoit\inst{\ref{neel}}
\and M. B$\rm\acute{e}$thermin\inst{\ref{stg}}
\and J. Bounmy\inst{\ref{lpsc}}
\and O. Bourrion\inst{\ref{lpsc}}
\and G. Bres\inst{\ref{neel}}
\and C. De Breuck\inst{\ref{eso_germany}}
\and M. Calvo\inst{\ref{neel}}
\and A. Catalano\inst{\ref{lpsc}}
\and C. Dubois\inst{\ref{lam}}
\and C.A Dur$\rm\acute{a}$n\inst{\ref{eso_germany}, \ref{IRAME}}
\and T. Fenouillet\inst{\ref{lam}}
\and J. Garcia\inst{\ref{lam}}
\and G. Garde\inst{\ref{neel}}
\and J. Goupy\inst{\ref{neel}}
\and C. Hoarau\inst{\ref{lpsc}}
\and W. Hu\inst{\ref{uwc}} 
\and J.-C. Lambert\inst{\ref{lam}}
\and F. Levy-Bertrand\inst{\ref{neel}}
\and J. Marpaud\inst{\ref{lpsc}}
\and R. Parra\inst{\ref{eso_chile}}
\and G. Pisano\inst{\ref{aig}}
\and N. Ponthieu\inst{\ref{ipag}}
\and L. Prieur\inst{\ref{lam}}
\and D. Quinatoa\inst{\ref{equ_Daysi}}
\and S. Roni\inst{\ref{lpsc}}
\and S. Roudier\inst{\ref{lpsc}}
\and D. Tourres\inst{\ref{lpsc}}
\and C. Tucker\inst{\ref{aig}}
\and M. Van Cuyck\inst{\ref{MVC_US}} 
}

\institute{Aix Marseille Univ, CNRS, CNES, LAM, Marseille, France, \label{lam} \email{andreas.lundgren@lam.fr}
    \and
            Univ. Grenoble Alpes, CNRS, IPAG, 38400 Saint Martin d’Héres, France\label{ipag}
        \and 
            Instituto de Astrofísica de Canarias, E-38205 La Laguna, Tenerife, Spain\label{iac}
        \and
            Departamento de Astrofísica, Universidad de La Laguna (ULL), E-38206 La Laguna, Tenerife, Spain\label{ull}       
        \and
            Univ. Grenoble Alpes, CNRS, Grenoble INP, LPSC-IN2P3, 53, avenue des Martyrs, 38000 Grenoble, France\label{lpsc} 
        \and 
            Astronomy Instrumentation Group, University of Cardiff, The Parade, CF24 3AA, United Kingdom\label{aig}
        \and
            Instituto de Estudios Astrof\'{\i}sicos, Facultad de Ingenier\'{\i}a y Ciencias, Universidad Diego Portales, Av. Ej\'ercito 441, Santiago, Chile\label{Santiago}   
        \and
            Univ. Grenoble Alpes, CNRS, Grenoble INP, Institut Néel, 38000 Grenoble, France\label{neel}
        \and
             Université de Strasbourg, CNRS, Observatoire astronomique de Strasbourg, UMR 7550, 67000 Strasbourg, France\label{stg}
        \and
            European Southern Observatory, Karl Schwarzschild Straße 2, 85748 Garching, Germany\label{eso_germany}
        \and    
            Instituto de Radioastronom\'ia Milim\'etrica (IRAM), Granada, Spain\label{IRAME}
        \and
            Department of Physics and Astronomy, University of the Western Cape, Robert Sobukhwe Road, Bellville, 7535, South Africa\label{uwc} 
        \and
            ESO Vitacura, Alonso de Córdova 3107, Vitacura, Casilla 19001, Santiago de Chile, Chile\label{eso_chile}
        \and
            Observatorio Astronómico de Quito, Escuela Politécnica Nacional, Quito 170403, Ecuador\label{equ_Daysi} 
        \and
        Department of Astronomy, University of Illinois, 1002 West Green Street, Urbana, IL 61801, USA \label{MVC_US} 
            }

   \date{Received X X, XXXX; accepted X X, XXXX}

  \abstract
  {The CarbON [CII] line in post-rEionisation and ReionisaTiOn epoch (CONCERTO) instrument is a low-resolution mapping  Fourier-transform spectrometer, based on lumped-element kinetic inductance detector (LEKID) technology, operating at 130--310\,GHz. It was installed on the 12-meter APEX telescope in Chile in April 2021 and operated until December 2022. CONCERTO's main science goal is to constrain the [CII] line fluctuations at high redshift. To reach that goal CONCERTO observed 1.4 deg$^2$ in the COSMOS field.}
{To ensure accurate calibration of the data, we have developed a forward model capable of simulating both the spectral response and the corresponding interferograms for each scan of observation in the COSMOS field.
   We present the modelling approach that enables us to reproduce the expected instrument outputs under controlled input conditions and provides a framework for the different calibration steps, including the absolute brightness calibration of the spectra.}
  {We constructed a dedicated analysis pipeline to characterise the raw interferometric data (interferograms) obtained under a broad range of atmospheric conditions at APEX. Using the forward model, we measured the interferogram alignment with the optical path difference (zero path difference, ZPD) and the relative response of each KID (flatfield). 
  Together, these elements enable a robust characterisation of the instrument’s spectral brightness calibration.} 
  {We demonstrate that the zero path difference systematically varies with elevation and across detectors, with variations that are consistent with small optical misalignments and elevation-dependent mechanical effects in the optical structure. The full measurement of these variations allow us to construct a data base that is used to accurately determine the zero path difference for each measured individual interferogram. The flatfield shows systematic variations with detector position but is extremely stable with time and atmospheric contribution. The accurate determination of the zero path differences and flatfields allows us to construct spectral cubes that combine all detectors and all blocks of data. Finally we present a novel method to calibrate the absolute brightness of those spectral cubes, which is immune to the exact knowledge of the bandpasses and directly applicable to extended emission.}
  {Our analysis establishes a framework for precise calibration directly from on-sky data. This approach ensures reliable performance for cosmological and astrophysical applications and can be readily adapted to future Martin–Puplett interferometer–based Fourier-transform spectrometers.}

   \maketitle
    \nolinenumbers

\section{Introduction}
Modern astrophysics increasingly relies on instruments of high complexity, designed to address demanding scientific objectives with unprecedented accuracy. Meeting these objectives requires not only advances in hardware but also a detailed understanding of instrumental behaviour. Comprehensive modelling has thus become indispensable: it enables the prediction of performance, the identification and mitigation of systematic effects, and the validation of data analysis pipelines. 

In the (sub-)millimetre, Fourier Transform Spectrometer (FTS) mapping meets the requirements for wide-area coverage and low-resolution spectroscopy in astronomical observations.
By recording interferograms generated from the superposition of two optical beams and applying Fourier analysis, FTS delivers spectral information with broad frequency coverage, modest spectral resolution (with  R$\sim$100-500), and full field of view (FoV) sampling in a single measurement. 
This combination has proven particularly valuable in studies of the Cosmic Microwave Background (CMB), as exemplified by the FIRAS instrument (\citealt{FIRAS}) and other space-borne instruments such as SPIRE (\citealt{2010A&A...518L...3G}). It has also been adopted as a baseline for several proposed above-the-atmosphere experiments, including PIXIE (\citealt{2025JCAP...04..020K}) and BISOU (\citealt{2024SPIE13102E}). For ground-based observations, atmospheric fluctuations require the use of fast detectors, with kinetic inductance detectors (KIDs) representing a well suited technology  available in large-format arrays at mm wavelengths (\citealt{2020A&A...641A.179C}). KIDs have already been deployed in photometric mm-instruments such as NIKA (\citealt{monfardini2010}) and NIKA2 (\citealt{adam2018}), which were both installed at the IRAM 30-m telescope at Pico Veleta (and in the case for NIKA2, still is). A particularly well-suited implementation of FTS in this context is the Martin–Puplett Interferometer (MPI, \citealt{mpi}), which has already been employed in experiments such as KISS (\citealt{fasano-ltd}) and OLIMPO (\citealt{Masi_2008_OLIMPO}), and constitutes the core design of the CONCERTO spectrometer.

The CONCERTO project (\citealt{2021NatAs...5..970M}, \citealt{2020A&A...642A..60C}) was conceived as a pathfinder to demonstrate wide-field spectroscopic mapping of the mm sky using an MPI-based FTS coupled to KID arrays. 

To maximise its scientific potential, CONCERTO relies on modelling approaches at different levels of detail, each designed to capture specific aspects of the instrument behaviour. 
In this work, we focus on a forward model of raw CONCERTO interferograms, designed to describe in detail how instrumental, optical, and environmental effects propagate into the measured interferometric signal. This interferogram-level modelling is a key ingredient for calibration, for the identification of systematic effects, and for the interpretation of the resulting spectra.

Higher-level, end-to-end modelling aimed at reproducing global instrument performance and observational characteristics is beyond the scope of this work. 
Here, we describe the forward modelling and characterisation of the raw interferograms measured by CONCERTO, together with the correction procedures required to reconstruct accurate spectra and derive their absolute calibration in astrophysical units.

The CONCERTO instrument was installed on the APEX telescope \citep{2006A&A...454L..13G} in April 2021. Commissioning runs were conducted shortly thereafter, and regular science operations took place between July 2021 and December 2022. 
The main scientific programme is a line-intensity mapping survey of the COSMOS field \citep{2007ApJS..172....1S}, complemented by dedicated observations of the Sunyaev–Zeldovich effect in galaxy clusters, star-forming regions (including the Galactic Centre), and evolved stars, all carried out during open time.
For a comprehensive description of the observing campaign and operational procedures, we refer the reader to \citet{hu2024}. The first scientific exploitation of CONCERTO data has recently been reported, with observations of the Orion Nebula revealing both CO(2–1) and water emission lines alongside the continuum. These results showcase the instrument’s spectral capabilities; see \citet{desert2025} for a detailed discussion.

The paper is organised as follows. Section\,\ref{sect:instrum} presents briefly the CONCERTO instrument. Section\,\ref{sect:obs} describes the observations we used in our analysis.
In Sect.\,\ref{sect:data_red}, we provide a summary of the data reduction that is needed for our work.
Section\,\ref{sect:fomo} develops the forward data model of CONCERTO raw interferograms, accounting for atmospheric emission, reference-source and stray-light contributions.
In Section\,\ref{sect:zpd_ff}, we report on the measurements of the zero path difference and the instrumental response from interferogram timelines for each detector. In Sect.\,\ref{sect:spectra}, we focus on an original method to derive the absolute brightness calibration of spectra. Finally, Sect.\,\ref{sect:cl} summarises our conclusions and discusses prospects for future applications.

\section{The instrument \label{sect:instrum}}

\begin{figure}[ht]
\centering
\includegraphics[width=.45\textwidth]{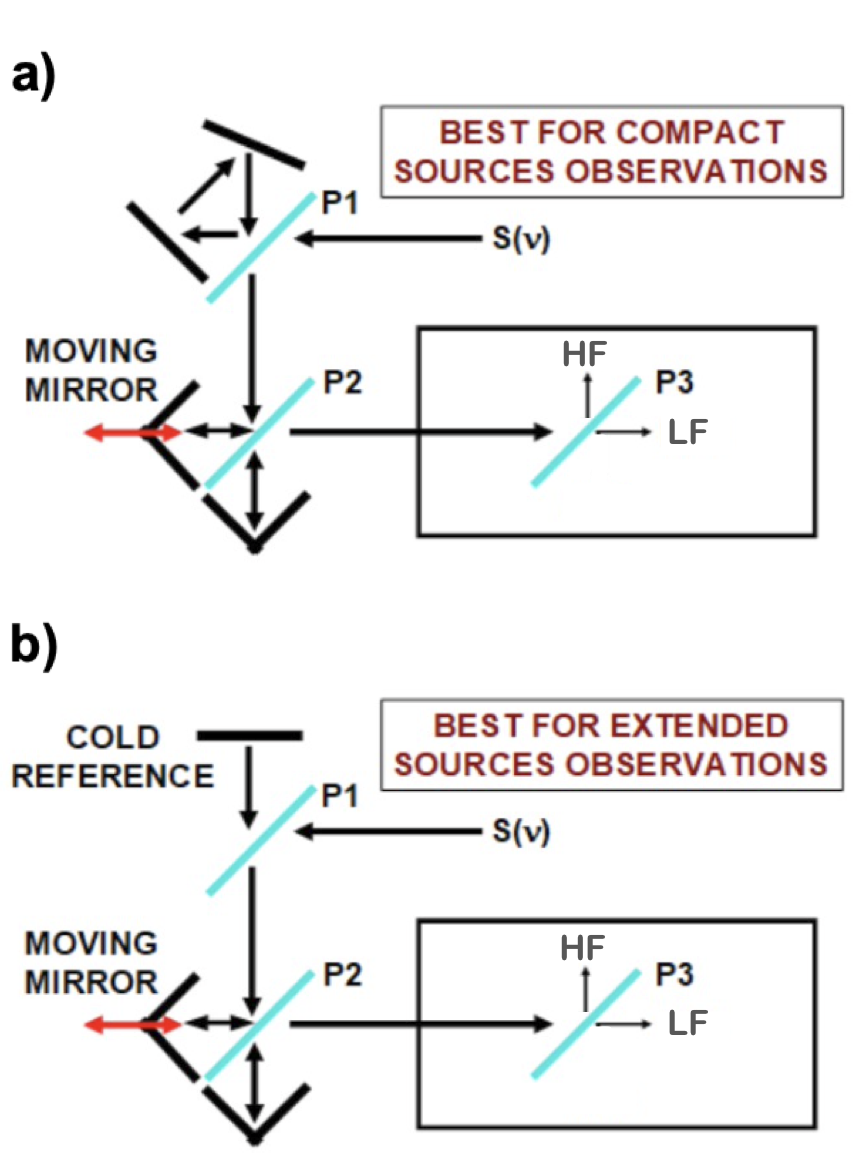}
\caption{Schematic view of the CONCERTO spectrometer. The incoming beam, represented by its spectral distribution $S_\nu$, is directed through the Martin–Puplett Interferometer. Two options are implemented for the reference input: (a) a defocused image of the $\sim18.6^\prime$ instantaneous field of view (REFSKY), and (b) a stabilised cold blackbody source (REFBB). 
Polarisers P1 and P2 are located at ambient pressure, while polariser P3 is housed in a cryostat together with the LF and HF detector arrays.
P1 provides the polarised input to the MPI, P2 acts as the beam splitter defining the two interferometric arms, and P3 defines the two complementary outputs of the interferometer, which are directed to the two focal-plane arrays.
The transmission output corresponds to the low-frequency (LF; 130–270\,GHz) band, and the reflection output to the high-frequency (HF; 195–310\,GHz) band.}
\label{fig:schematic}
\end{figure}

The CONCERTO spectrometer employs a Martin–Puplett Interferometer, a configuration also used in its precursor, the KIDs Interferometric Spectral Surveyor (KISS, \citealt{fasano-ltd}, \citealt{2024PASP..136k4505M}). The MPI combines two input sources and introduces an optical path difference (OPD) through the displacement of a rooftop mirror, producing an interferogram recorded at two complementary outputs, as illustrated in Fig.~\ref{fig:schematic}. By adjusting the stroke of the moving mirror, the OPD can be tuned, with longer strokes providing higher spectral resolution.

For clarity, polarisers P1 and P2 are located at ambient pressure and temperature within the instrument enclosure, while polariser P3 is located inside the cryostat together with the LF and HF detector arrays, which operate under vacuum at cryogenic temperatures.

In the Martin-Puplett configuration, the interferometer intrinsically measures the differential signal between the sky-facing input and a reference input.
In CONCERTO, two reference configurations are implemented:
\begin{itemize}
 \item{REFSKY: a defocused image of the $\sim18.6$ \,arcmin instantaneous field of view, which optically subtracts the atmospheric common-mode spectrum and is well suited for compact sources, smaller than the FoV \citep{2020A&A...642A..60C,desert2025}}.
 \item{REFBB: a stabilised cold blackbody source, which serves as the reference for the present work (see Sect.~\ref{subsec:refsource} for details)}. 
\end{itemize}
Since this article exclusively addresses COSMOS data acquired in REFBB mode, our discussion will focus on this.

The transmission output corresponds to the low-frequency band (LF; 130–270\,GHz), while the reflection output corresponds to the high-frequency band (HF; 195–310\,GHz). A schematised view of the core of the instrument is presented in Fig.~\ref{fig:schematic}, highlighting key components such as the polariser P3 that splits the signal into the  corresponding LF and HF focal planes. 

The LF and HF arrays are large-format assemblies of lumped-element kinetic inductance detectors (LEKIDs), superconducting resonators operated at cryogenic temperatures. A central advantage of LEKIDs in Fourier transform spectroscopy is their ability to linearly record the large signal variations expected for interferograms of atmospheric emission. 
In these detectors, incident radiation modifies the kinetic inductance of the superconducting film, leading to a shift in the resonance frequency of each KID, which is monitored by the readout electronics. In addition, the intrinsic fast time response of LEKIDs makes them particularly well suited to the rapid sampling required to mitigate atmospheric instabilities and fast-varying interferograms \citep{2020A&A...641A.179C}. 
The main source of systematic error may instead arise from the measurement of these resonance-frequency shifts, particularly when rapid variations in the background radiation move the detector resonance far from the fixed readout tone frequency.

Each frequency band is equipped with 2152 LEKIDs \citep{2010SPIE.7741E..0MD}. The detectors are coupled to planar antennas and they are read out using frequency-division multiplexing, which allows 400 resonators to be monitored simultaneously over a single transmission line (\citealt{2022JInst..17P0047B}).

\begin{figure*}[ht]
\centering
\includegraphics[width=.95\textwidth]{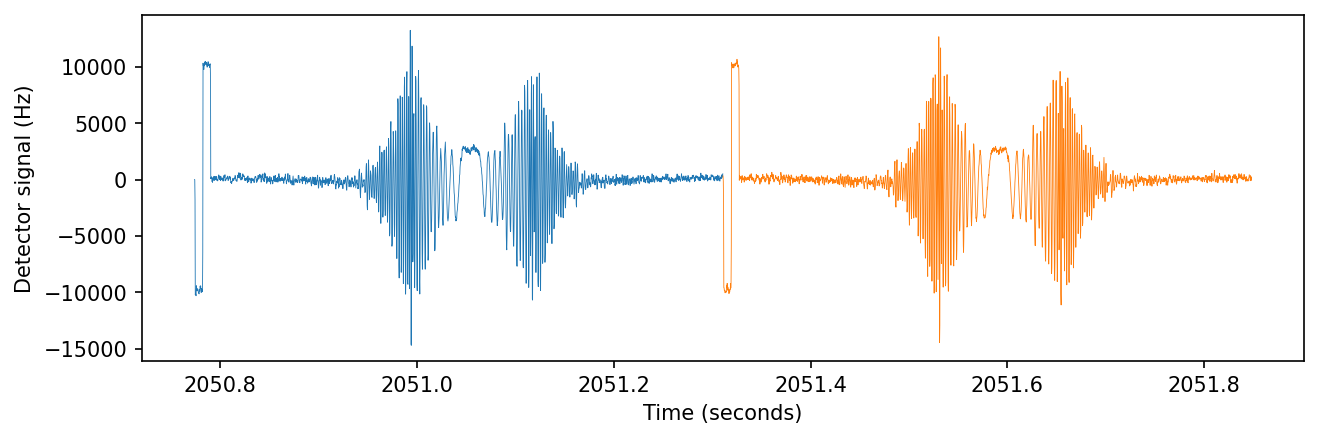}
\caption{
Two representative interferometric blocks from real CONCERTO observations for one KID, with time on the x-axis and detector signal (Hz) on the y-axis. Each block (shown in blue and orange) corresponds to a full rooftop mirror roundtrip (forward and backward) scan, defining the optical path difference (OPD) and setting the effective spectral resolution. The first squared-shaped samples correspond to the modulation used for the 3-point calibration (\citealt{2022JInst..17P8037B}). They are followed by the forward and then the backward interferogram.
}
\label{fig:block}
\end{figure*}

The practical implementation for one KID is illustrated in Fig.~\ref{fig:block}, which shows two representative interferometric blocks from COSMOS-field observations. Each block corresponds to a full cycle of rooftop mirror movement, defining the OPD and setting the effective spectral resolution. This block-level organisation ensures that raw measurements can be efficiently transformed into calibrated interferograms and astrophysical spectra. A more detailed description of the acquisition chain, including mirror motion, sampling strategy, and KID readout, is presented in \citet{2022JInst..17P8037B}, \citet{2022JInst..17P0047B} and \citet{desert2025}.

\section{Observations \label{sect:obs}}

In this work, we use a subset of COSMOS field observations obtained as part of the CONCERTO [CII] Line Intensity Mapping (LIM) Large Programme. Between July 2021 and December 2022, the project accumulated 793 hours on the field, combining rectangular and spiral on-the-fly (OTF) raster scans for a total of 1522 individual scans. These observations were acquired with REFBB as the reference source and span a wide range of atmospheric conditions (PWV = 0.3–3 mm) and source elevations (24$^\circ$–65$^\circ$).

Each rectangular or spiral OTF scan corresponds to the mapping of a predefined RA/Dec region of the COSMOS field, with a typical scan duration of 20–25 minutes. During an individual scan, the elevation changes by approximately 1–8$^\circ$, with the smallest variation occurring near transit and the largest variation at low elevations. Over the course of an observing session, the field is tracked from low elevation through transit, reaching a maximum elevation of about 65$^\circ$.

For the present analysis, we exclusively use data extracted from individual scans, selecting as many contiguous subsets of 50 interferogram blocks (“slices”) as possible within a given scan. This approach limits the elevation range sampled within each subset, which is well suited to the forward-modelling and calibration analyses presented in this work, while maintaining computational efficiency. 
The LEKIDs were operated with fixed tuning parameters during each scan and were not re-tuned in response to elevation-dependent background variations within a scan \citep{2022JInst..17P8037B}. However, the detectors were re-tuned between scans to maintain optimal operating conditions.

\section{Data reduction \label{sect:data_red}}
The spectroscopic data reduction pipeline with REFSKY reference was partially described in \citet{desert2025} and a complete description of both reduction pipelines will be presented in Beelen el al.\ (in prep.). 
We distinguish between the time-ordered data (TOD), corresponding to the raw detector readout as a function of time, and the interferogram timelines, which are obtained after basic data reduction and scan reconstruction by mapping the TOD onto optical path difference. The interferogram timelines therefore represent the reduced data products used as input to the forward modelling, rather than the original detector time streams.
We summarise here the only two steps of the data reduction process that are relevant for the analysis presented in this work:

\begin{itemize}

\item (I,Q) to Hz: The raw time-ordered data (TOD) are recorded as in-phase (I) and quadrature (Q) components of the complex KID readout signal (not to be confused with Stokes parameters), corresponding to the mixing between the excitation tone and the detector output. The data are organised into blocks of 2048 samples per detector, sampled at 3814.7\,Hz (500 MHz/2$^{17}$, where 500 MHz is the original clock frequency). The conversion to a resonance frequency shift $\Delta f$ in Hz is performed independently for each block (Fig.\,\ref{fig:block}) using the 3-point modulation algorithm \citep{fasano2021, 2022JInst..17P8037B}, in which the excitation tone is rapidly modulated between two known calibration frequencies and a science frequency corresponding to the tuned resonance. This scheme reconstructs the resonator (I,Q) circle and converts the measured phase signal into a resonance-frequency shift, thereby reducing off-resonance systematic errors.
In KID detectors, incident optical power modifies the kinetic inductance of the resonator, leading to a shift of its resonance frequency; this frequency shift therefore constitutes the detector response and explains why interferograms are expressed in units of Hz. The resulting $\Delta f$ provides a calibrated measure of the optical load.

\item KID selection: a careful identification and removal of non-optimal detectors before calibration and analysis is mandatory. As detailed in \cite{hu2024}, the selection process is based on a combination of stability, responsivity, and noise performance. Detectors exhibiting pathological behaviour—such as unstable resonance frequency, anomalous gain variations, or strong 1/f noise components—are flagged and excluded from subsequent processing. Additional rejection criteria are applied to detectors with persistent electronic cross-talk or with response patterns inconsistent with the optical beam.

\end{itemize}

\section{Modelling the raw interferograms of CONCERTO \label{sect:fomo}}

The modelling of CONCERTO’s instrumental response is naturally organised into two complementary steps. In spectral space, we describe the astrophysical and atmospheric signals together with the instrumental transmission, producing a compound spectrum.
This spectrum is then transformed into Fourier space, yielding model interferograms that can be directly compared with the raw measurements. 

The forward model is restricted to the dominant contributions that determine the large-scale structure of the interferograms. 
In the present implementation we include atmospheric emission, the reference source, and a stray-light component, all filtered by the measured instrument bandpasses. 
Astrophysical sky emission is neglected because its contribution is small compared to the atmospheric and instrumental backgrounds for the calibration observations analysed here. 
We further assume that the detector response is linear over the range of optical loading encountered in the data and that instrumental parameters such as the bandpass and stray-light contribution remain constant within a given scan. 
Under these assumptions the model captures the leading-order behaviour of the interferograms while remaining computationally tractable for the large number of detectors and scans considered in this work.

In this way the modelling follows the same flow as the data: the physical signals are naturally described in frequency space, while the instrument records them as interferograms.

\subsection{Description in spectrum space}

To interpret the measured interferograms, it is necessary to construct forward models of the signals entering the two beams of the spectrometer. On the one side, the model accounts for the atmospheric emission, which varies with PWV and telescope elevation and must therefore be updated for each observation.
The contributions of the reference source (REFBB) and stray light are kept fixed with respect to elevation and PWV in the present model. 

The models are combined with the measured instrument bandpass to form a compound spectrum, which is then Fourier transformed into interferogram space. 
In this formulation the bandpass is assumed to be stable during a scan, allowing the instrumental response to be represented by a fixed spectral weighting applied to the incoming radiation.
This procedure ensures proper sampling of the interferograms, with the frequency range and spacing in spectrum space chosen to guarantee accuracy in Fourier space. 

In the following subsections, we describe in more detail the three main ingredients of this modelling: the atmosphere, the reference source, and stray light.

\subsubsection{Atmospheric contribution}

The atmosphere is a major contributor to the CONCERTO signal, with its relative importance depending on frequency band and observing conditions. Away from strong atmospheric emission features, including the 183\, GHz water vapour line, and at low PWV, the spectral atmospheric emission in the lower frequency range is typically lower than the contribution from the reference source. In contrast, in the upper frequency range, or at higher PWV, the atmospheric continuum becomes more dominant due to the increased radiative background.

To represent the atmospheric transmission and emission, the forward model makes use of the ATM code \citep{pardo2001, pardo2025},
which combines spectroscopic line catalogs with a semi-empirical treatment of water vapour continuum absorption. 
The model spectra exhibit the characteristic absorption features of water vapour and molecular oxygen, which are imprinted on the observed data and provide valuable anchors for validating the instrument’s frequency scale. In addition, the broadband level of atmospheric emission sets a significant fraction of the optical loading on the detectors, directly impacting sensitivity. 

In order to model the atmosphere accurately while keeping the processing time manageable, we employ a pre-computed lookup table. For each observation, the atmospheric spectrum is retrieved from this table based on the PWV and telescope azimuth and elevation measured at the time of the scan. This strategy captures the main variations in atmospheric emission without requiring a full ATM run for every scan, and dense spacing between grid points ensures smooth coverage across parameter space. 

\subsubsection{Reference source contribution}
\label{subsec:refsource}

In the case of REFBB observations, the second input beam of the instrument is fed by a dedicated internal reference source, which provides a stable broadband spectrum against which the astronomical signal is measured. Its contribution is assumed to be constant in time and independent of observing conditions. The REFBB configuration is realised by a cryogenically cooled copper plate maintained at $\sim$20\,K using a pulse-tube cooler. The optical path of the reference source includes additional elements internal to the REFBB assembly: a 20 mm thick polypropylene window and two 0.25 mm thick Zitex layers thermally anchored at 50 K and 10 K, respectively, which serve to suppress infrared radiation. These elements are specific to the reference-source optical chain and are distinct from the sky-facing entrance optics of the CONCERTO camera \citep{fasano2024}.

When modelled with the Planck law, these components yield an effective optical load temperature that increases smoothly from 43\,K to 51\,K over the frequency range 110–350\,GHz, as characterised in \citet{fasano2024}. 
Although this effective load is characterised over a broad frequency range (110–350\,GHz), the band-limited reference contribution relevant for the LF and HF channels is obtained naturally in our forward model by weighting this spectrum with the measured instrument bandpasses.

\subsubsection{Stray light contribution}
\label{subsec:slc}

In addition to the atmosphere and reference source, a small fraction of the detected signal originates from stray light within the optical paths. This includes thermal emission from warm surfaces of the instrument and residual pickup from outside the nominal beams. 
It was modelled as a Cassegrain receiver cabin (C-cabin) ambient-temperature blackbody, at 11$^\circ$C \citep{fasano2024}.
Although subdominant, accounting for stray light ensures a more accurate representation of the interferogram baseline.

The net stray light, resulting from the difference between the two optical paths, has been detected and quantified using both skydip and COSMOS observations. 
The observed amplitudes are in line with expectations from the instrument design and construction. 
In addition, a systematic geometric pattern of stray light across the focal plane is observed, with second-order variations correlated with telescope elevation. This behaviour is likely linked to slow deformations of the beamsplitter projected onto the ground, which vary with elevation angle.

Following these results, we model the stray light as an empirically determined effective blackbody contribution originating from the C-cabin environment, corresponding to an optical load of $\sim$28\,K. This value reflects a net stray-light coupling factor of order 0.1 relative to the physical cabin temperature (284\,K), and was derived from two  approaches that agree within 1$\sigma$: fits to averaged COSMOS interferograms including a stray-light term, and an independent skydip analysis at zero optical path difference. We adopt this effective load in the present work as a first-order representation of the net stray-light contribution.

Including the contribution from REFBB (see Sect.\,\ref{subsec:refsource}), the total effective optical load increases to 71--78\,K over the frequency range 110-350\,GHz, which is subsequently band-limited by the instrument response for each channel. This reference load provides a stable and well-characterised baseline against which the atmospheric and astrophysical signals are measured.

\subsubsection{Compound spectra}
\label{sec:compound}
In addition to the individual signal components described above, the instrumental bandpass must be taken into account to combine them into a physically meaningful model spectrum. 
The bandpasses of the LF and HF arrays are used to combine the modelled atmospheric emission with the instrumental contributions (reference source and stray light) into forward-modelled interferograms, enabling direct comparison with the measured data.
We adopt these bandpasses from \citet{hu2024}, where they were measured by closing the APEX shutter and assuming blackbody illumination. 
This yields the relative spectral response appropriate for point sources or, equivalently, for Rayleigh--Jeans temperatures (see their Appendices B and C).

The total instrumental signal is modelled as the combination of the three main contributions (atmosphere, REFBB, and stray light). In the model, the atmospheric component is updated dynamically according to the contemporaneous PWV and elevation at the block level, in contrast to the reference source and stray light contributions, which remain fixed.
The sum of these three spectra is then weighted by the instrument bandpass (see Sect.\,\ref{sect:data_red}) to form a compound spectrum, which is then Fourier-transformed to generate the interferogram. The approach allows us to predict the expected shapes of interferograms under varying observing conditions. 

For a given measured interferogram of a detector $i$, the reconstructed spectrum of the input signal can be written as:
\begin{equation}\label{eq:miv}
m_{i,\nu} = k_i \bandpass \left[ e^{\amtn} \tcos + (1 - e^{\amtn}) \tatm - \tref - \tstr \right]
\end{equation}
where $\nu$ is the frequency, $k_i$ is the detector-dependent calibration factor [Hz/K], 
$\bandpass$ is the area normalised bandpass [GHz$^{-1}$], $AM = 1/\sin(el)$ the airmass at elevation $el$, $\tau_\nu$ the zenith opacity. The brightness temperatures $\tatm$, $\tref$, and $\tstr$ represent the atmosphere, the cold reference, and stray light contributions, respectively, while $\tcos$ is the astrophysical signal. 

The quantity $m_{i,\nu}$ represents a bandpass-weighted spectral contribution per unit frequency and is therefore expressed in units of Hz/GHz, where the numerator (Hz) reflects the change in the KIDs resonance frequency and the denominator (GHz) the spectral frequency.

Writing the measurement in temperature units ensures that the same bandpass function applies as for point sources \citep{hu2024}.

Since $\tcos$ is typically extremely small compared to the other contributions, it can thus be neglected in the rest of this study:
\begin{equation}\label{eq:epsmethod}
    m_{i,\nu} = k_i \bandpass \left[ \epsilon_\nu \tatm - \tref - \tstr \right] \,,
\end{equation}
where $\epsilon_\nu = 1 - e^{\amtn}$ is the atmospheric emissivity. Note that the ATM model is directly giving $\epsilon_\nu \tatm$ given the proper airmass and PWV content.

\subsection{Description in interferogram space}
While the astrophysical modelling is most naturally formulated in spectrum space, the instrument records data in interferogram space. Our forward data model must therefore transform the compound spectrum into the Fourier domain and evaluate it as a function of optical path difference (OPD), which is determined by the stroke of the moving mirror of the interferometer.

We denote the optical path difference by $\delta$, which is set by the position of the moving mirror.

For each interferogram, the OPD is reconstructed from the internal laser metrology system. CONCERTO employs three metrology lasers to monitor different optical elements of the interferometer. To determine the optical path difference $\delta$, we use the readout of the laser monitoring the moving rooftop mirror ($laser1$) together with that of a second laser ($laser3$) directed towards the P2 polariser membrane, which tracks residual vibrations after real-time correction. 
Earlier publications describe different subsets of this metrology system depending on the subsystem under study \citep[e.g.][]{fasano2022}.
Their combined signal is defined as $laser = laser1 + laser3$, and the OPD is then computed as

\begin{equation}\label{eq:opd}
\delta = 2 \times (\textit{laser} - \rm{ZPD})\,,
\end{equation}
where the factor of two accounts for the double pass through the interferometer. The zero path difference (ZPD) corresponds to the position where the optical paths of the two arms of the MPI are equal. It will be determined in Sect.\,\ref{sect:zpd_ff}.

\subsection{From spectrum space to interferogram space}
\label{sec:inv_spectrum_to_fourier}

For comparison with the interferometric data, the compound spectrum described in Sect.~\ref{sec:compound} is transformed into the Fourier domain to generate the corresponding model interferograms.

In CONCERTO, this transformation is performed numerically using a discrete Fast Fourier Transform (FFT). Specifically, we apply a real-to-complex FFT 
to the model spectrum array, yielding the corresponding complex interferogram. We use the highest spectral resolution input spectrum model to define a spectral binning on which all input spectra and bandpass are linearly interpolated. This defines a maximum OPD which is much higher than the one available with the mirror course of CONCERTO. By padding the input spectra and bandpass, we also produce interferograms with an higher OPD sampling to resolve all physical features.
This provides an oversampled model interferogram $I(\delta)$ as a function of optical path difference $\delta$ on a dense OPD grid. A linear interpolation is then constructed and evaluated at the measured OPD positions of each interferogram, thereby accounting for the slight irregular sampling of the mirror motion.

The interferograms derived from the compound spectra described above are fitted to the interferograms recorded by CONCERTO.
The fitting procedure uses as an initial guess a model value of the ZPD (derived in Sect.\,\ref{sec:zpd}). The fit yields refined estimates of the ZPD and the detector response $k_i$, which are stored in a database for subsequent use in the data reduction pipeline.
The workflow is summarised in Fig.\,\ref{fig:flow}. 

Accurate ZPD determination is essential, since even small offsets introduce phase errors in Fourier space, biasing the recovered spectra. For example, for a Gaussian line at 300\,GHz, a ZPD error of 0.035\,mm translates into a 10\% decrease in the recovered line amplitude. The determination of $k_i$ allows us to derive a spectroscopic flat-field, assuming that the LEKIDs all see the same signal.
Flat-fielding is particularly important for on-the-fly (OTF) observations and multi beam instruments, where the signal in each sky pixel is reconstructed from the combined contributions of many detectors.

\begin{figure}[ht]
    \centering
    \begin{tikzpicture}[
        node distance=0.4cm,
        box/.style={rectangle, draw, rounded corners, align=center, minimum width=4cm, minimum height=1cm},
        arrow/.style={-{Stealth[length=2mm]}, thick},
        input/.style={trapezium,trapezium stretches=true,draw,thick,minimum width=2.5cm,minimum height=1cm,trapezium left angle=70,trapezium right angle=110,align=center},
        data/.style={trapezium,trapezium stretches=true,draw,thick,fill=gray!10,minimum width=2.5cm,minimum height=1cm,trapezium left angle=110,trapezium right angle=70,align=center},
        process/.style={rectangle, rounded corners, draw, thick, fill=gray!30, align=center,  minimum height=1cm, minimum width=2.5cm},
        database/.style = {cylinder, cylinder uses custom fill, cylinder body fill=gray!10, cylinder end fill=gray!10, shape border rotate=90, aspect=0.25, draw}
    ]
    \node[input] (atmo) {Atmospheric \\ emission};
    \node[input, left=of atmo, xshift=-0.2cm] (refs) {Reference and \\ stray light};
    \node[input, right=of atmo, xshift=0.2cm] (bp) {Bandpass};
    \node[data, below=of atmo] (compound) {Compound spectrum};
    \node[process, below=of compound] (fourier) {Fourier transform \\ (\texttt{np.fft.rfft})};
    \node[data, below=of fourier] (interf) {Compound \\ interferogram};
    \node[input, left=of interf, xshift=-0.2cm] (zpdm) {ZPD model \\ initial guess};
    \node[input, right=of interf, xshift=0.2cm] (concerto) {CONCERTO \\ data};
    \node[process, below=of interf] (fit) {Parameter fitting: \\ ZPD \& Response};
    \node[database, below=of fit] (data) {ZPD and Response data base};
    \draw[arrow] (atmo) -- (compound);
    \draw[arrow] (refs) -- (compound);
    \draw[arrow] (bp) -- (compound);
    \draw[arrow] (compound) -- (fourier);
    \draw[arrow] (fourier) -- (interf);
    \draw[arrow] (interf) -- (fit);
    \draw[arrow] (zpdm) -- (fit);
    \draw[arrow] (concerto) -- (fit);
    \draw[arrow] (fit) -- (data);

    \end{tikzpicture}
    \caption{
        Flow diagram of the forward-model fitting procedure. Atmospheric emission, reference source, and stray light are combined and filtered by the instrument bandpass to form the compound spectrum, which is Fourier transformed into the compound interferogram. Using a model ZPD value as the initial guess (see Sect\,\ref{sec:zpd}), the interferogram is fitted to the data, yielding refined ZPD and response estimates that are stored in a database for later use.
    }
    \label{fig:flow}
\end{figure}

With two frequency channels (LF and HF), about 1300 selected LEKIDs per array, and roughly 2830 science blocks per scan, each detector records about 5660 interferograms (forward and backward), corresponding to approximately 14.7 million science interferograms per scan across the full instrument.
The fitting procedure treats each interferogram independently and in parallel, returning the two fitted parameters (ZPD and the detector response $k_i$).

\begin{figure*}[ht]
\centering
\includegraphics[width=.48\textwidth]{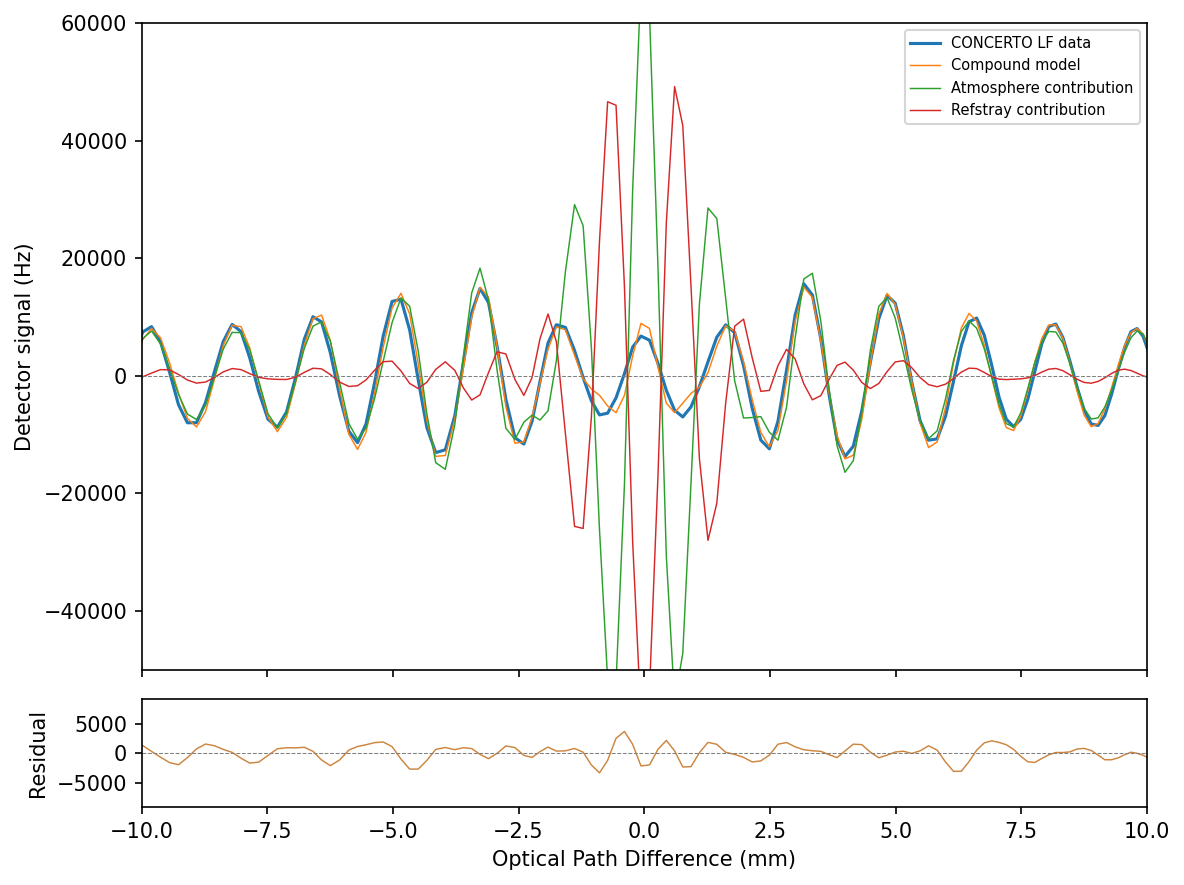}
\includegraphics[width=.48\textwidth]{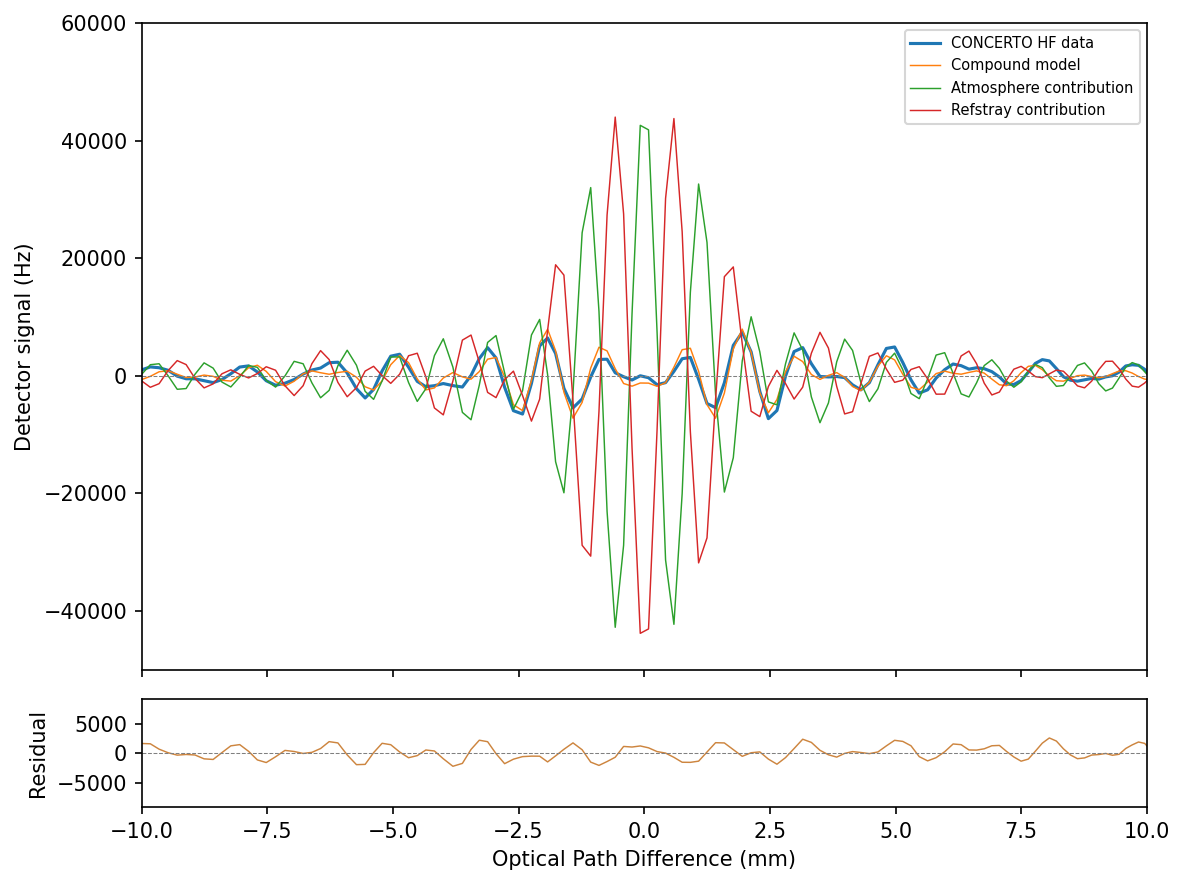}
\caption{The left panel shows an LF interferogram obtained at an elevation of 24$^\circ$ and PWV of 2.17\,mm. In this band, the strong water-vapour line at 183\,GHz produces a dominant atmospheric contribution (green), which is comparable in amplitude to the combined reference and stray-light emission (“refstray,” red). Their sum (orange) closely reproduces the observed data (blue), such that model and measurement nearly overlap. The right panel shows an HF interferogram at 41$^\circ$ elevation and PWV of 0.58\,mm. Here the absence of strong atmospheric lines and the lower PWV suppress atmospheric emission, leaving the refstray component as the main contributor. In both cases, the compound model tracks the data with high accuracy, with only small deviations visible away from the interferogram centre. 
The bottom panel shows the residuals (data minus model).
}
\label{fig:itg_fomo}
\end{figure*}

We show in Fig.\,\ref{fig:itg_fomo} the fitted interferogram models for the atmospheric contribution and the combined REFBB + stray light component (hereafter “refstray”) in both LF and HF channels, for two clearly different cases of elevation and PWV. In the LF case, observed at low elevation and high PWV, the strong 183\,GHz water-vapour line dominates the atmospheric signal, resulting in an interferogram where the atmospheric and refstray components contribute with comparable amplitudes. 
Their combination reproduces the measured data with remarkable accuracy, such that model and observation are nearly indistinguishable. In contrast, the HF case was obtained under drier conditions and at higher elevation, where the absence of strong atmospheric lines leaves the refstray emission as the main contributor to the interferogram. Here too, the compound model closely tracks the data. 

\section{Measuring the ZPD and the instrumental response from interferogram timelines \label{sect:zpd_ff}}

We present in this section the results obtained by comparing the forward model predictions with the measured interferograms and by optimising a limited set of instrumental calibration parameters, namely the ZPD (Sect.\,\ref{sec:zpd}) and the relative detector response (Sect.\,\ref{sec:resp}). The underlying approach remains a forward modelling of the interferometric signal from physical inputs; the fitting procedure is used solely to determine these instrumental parameters and does not involve any inversion of the astrophysical spectra.

\subsection{Zero path difference}
\label{sec:zpd}

For CONCERTO, the location of the ZPD varies across detectors and changes systematically with telescope elevation. These variations are likely related to small optical misalignments and elevation-dependent changes in the instrument geometry, although we do not attempt to model the underlying mechanical behaviour in this work.
To characterise this effect, we analysed a large number of COSMOS scans covering a wide range of PWVs and elevations. 

We used the forward model to fit the ZPD for interferograms extracted from multiple COSMOS scans and grouped into slices of 50 interferogram blocks (see Sect.~\ref{sect:obs}). From these fits we derived an empirical description of the ZPD dependence on both detector position in the field of view and telescope elevation.
This empirical model is then used to provide an initial estimate of the ZPD in the data reduction pipeline, where the ZPD is fitted for each block.

\begin{figure*}[ht]
\centering
\includegraphics[width=.48\textwidth]{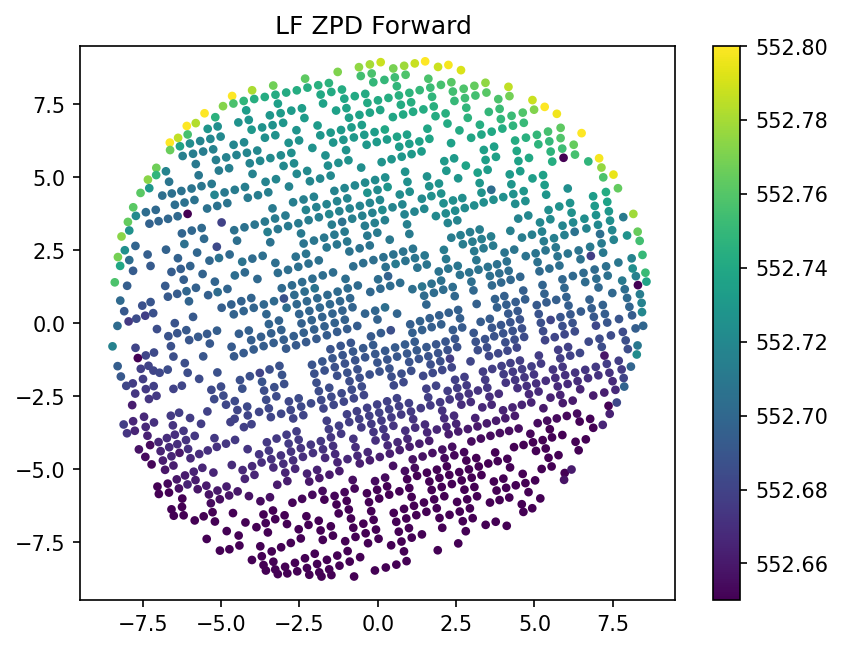}
\includegraphics[width=.48\textwidth]{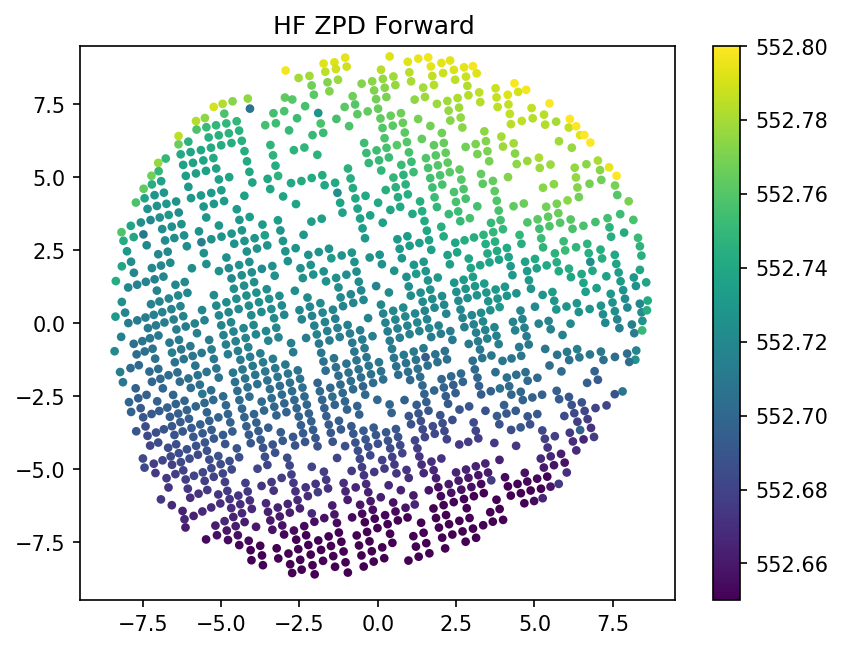}
\caption{
Median ZPD values (in mm) derived for individual KIDs in the LF (left) and HF (right) arrays. Values are shown for forward interferograms; the backward solutions are nearly, though not exactly, identical and therefore must be fitted separately. The LF and HF maps exhibit very similar spatial patterns, but differ in absolute ranges (indicated by the colour bars). The sequence of the six feed lines is visible in both arrays, delineated by the empty lines. For clarity, these spatial variations are shown independently of the elevation dependence, though in practice both effects are fitted simultaneously.}
\label{fig:ZPD_arrays}
\end{figure*}

Figure\,\ref{fig:ZPD_arrays} shows maps of the median ZPD for all valid detectors in the LF and HF arrays, derived from a COSMOS slice at 45$^\circ$. The two arrays exhibit broadly similar, though not identical, spatial patterns. Forward and backward interferograms yield nearly identical maps, but small systematic differences remain and therefore must be fitted separately.

\begin{figure*}[ht]
\centering
\includegraphics[width=.48\textwidth]{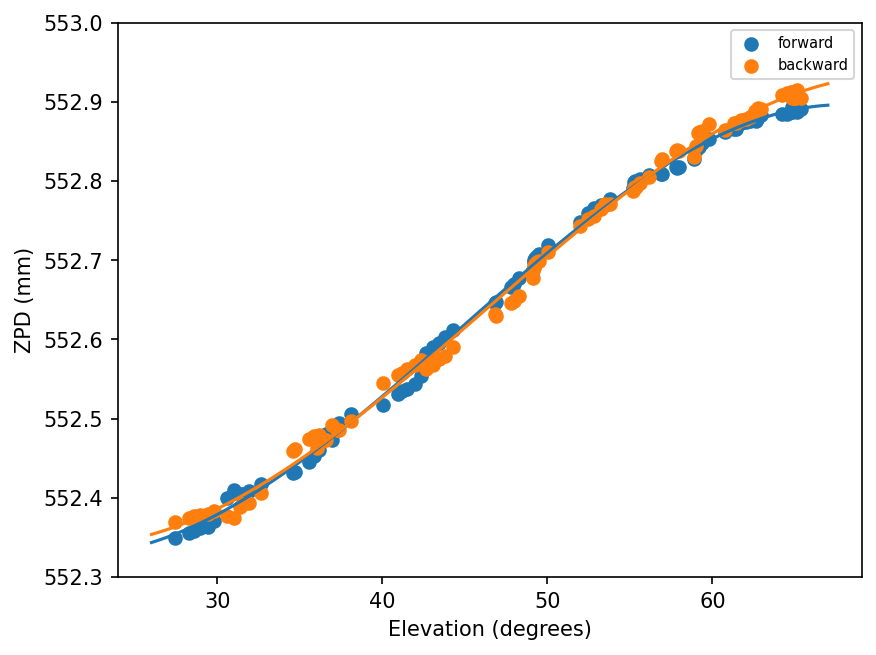}
\includegraphics[width=.48\textwidth]{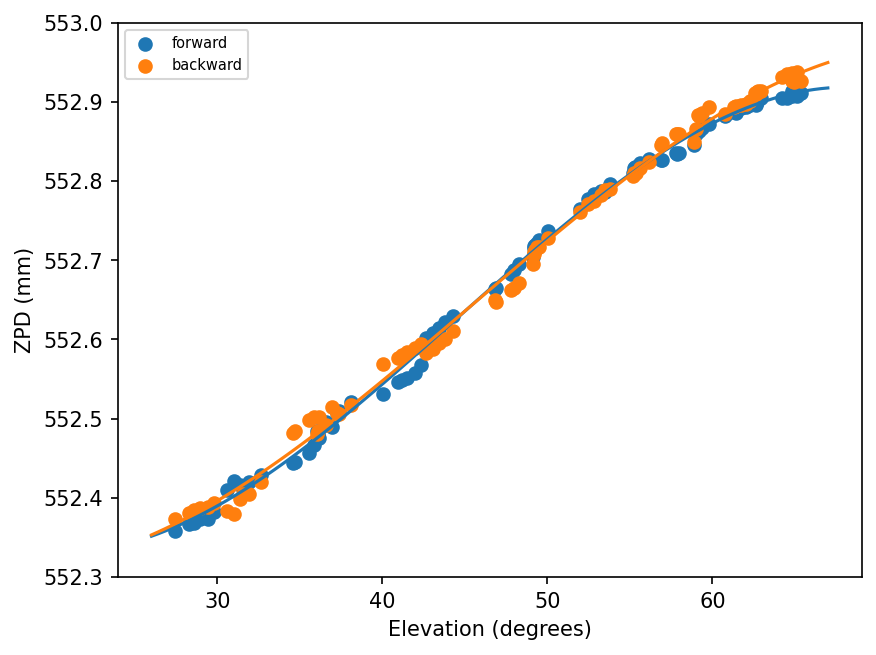}
\caption{
Median ZPD (mm) as a function of elevation for forward (blue) and backward (orange) interferograms. LF is shown on the left and HF on the right, together with fitted curves describing the elevation trends. These trends are illustrated separately from the position-dependent variations across the arrays (Fig.\,\ref{fig:ZPD_arrays}), but in practice both are modelled simultaneously to construct a three-dimensional empirical ZPD model. 
}
\label{fig:ZPD_elevations}
\end{figure*}

In Fig.\,\ref{fig:ZPD_elevations} we plot the median ZPD as a function of elevation for forward and backward interferograms. The fitted trends (third-order polynomials) capture the elevation dependence in both arrays. 
Over the elevation range sampled in this study (24$^\circ$–65$^\circ$), the ZPD shifts reach peak-to-peak amplitudes of up to approximately 0.5\,mm across the array.
While we illustrate the array, and elevation, dependent variations separately (Figs.\,\ref{fig:ZPD_arrays} and \ref{fig:ZPD_elevations}), in practice they are fitted simultaneously over the full elevation range. This yields a three-dimensional empirical model of the predicted ZPD, combining detector position in the array with telescope elevation. The resulting model provides the initial guess for ZPD in the CONCERTO fitting procedure. The accuracy of this guess is critical: it must be better than 0.2\,mm to prevent the fitting routine from locking onto a sidelobe instead of the true interferogram centre.

\subsection{Diffuse emission spectroscopic flat-field}
\label{sec:resp}

The relative response of each KID is determined from the same data set used to establish the ZPD model. For each detector, we measure the median interferometric response and normalise it to the global array, deriving an interferogram-based flat-field. This flat-field is appropriate for diffuse emission in spectroscopic mode as it was derived mainly from atmospheric emission. It differs from the main-beam continuum flat-field presented in \cite{hu2024} appropriate for point sources while CONCERTO is in pure continuum mode. 

After accounting for the spectroscopic efficiency of the MPI, this diffuse spectroscopic flat-field shows strong similarities with the point-source continuum flat-field once the latter is converted to diffuse emission using the individual LEKID beams. This indicates that the spectroscopic efficiency is relatively uniform across the array.

\begin{figure*}[ht]
\centering
\includegraphics[width=.48\textwidth]{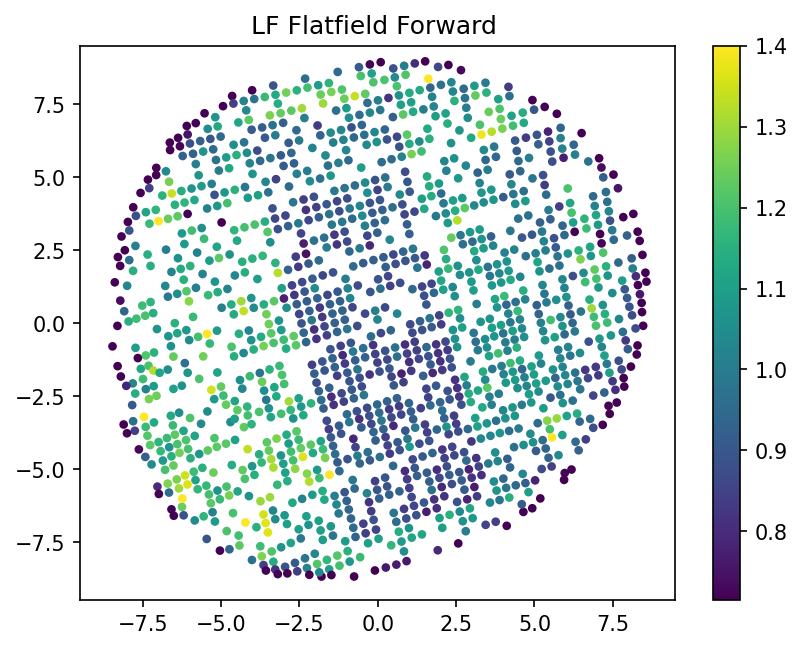}
\includegraphics[width=.48\textwidth]{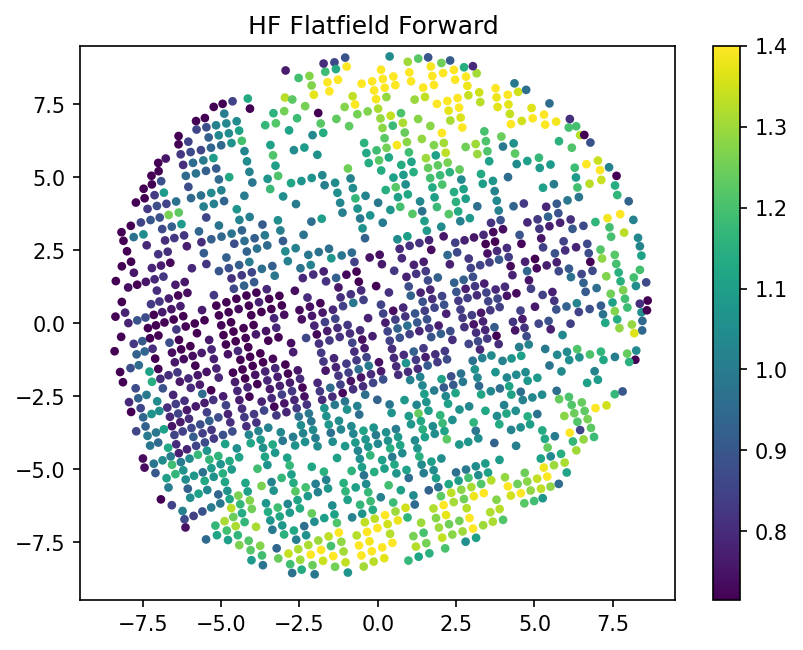}
\caption{Flat-fields derived from the interferogram response to the forward model. The left (right) figure shows the LF (HF) array. We display for each LEKID its average response value, normalized to the global array median.}
\label{fig:flat-fields}
\end{figure*}

In Fig.\,\ref{fig:flat-fields} we present the median normalised flat-field. Overall, the distribution of the flat-field shows a mild gradient perpendicular to the feed lines, with distinct zones reflecting the grouping of resonance frequencies of the LEKIDs along each line. These groups correspond to electronic sub-bands, which are amplified with different gains to compensate for the intrinsic increase in readout noise at higher resonance frequencies.

The derived flat-field is remarkably stable. Within a given observing run, flat-fields obtained from different elevation slices are nearly identical, with only very small dispersions. Comparing flat-fields across different elevations and atmospheric conditions yields a median absolute deviation of 2-3\% for both the LF and HF arrays. This level of variation is small compared to the overall detector-to-detector response differences and indicates that the relative pattern of detector responses is highly preserved. Such variations represent a minor contribution to the systematic uncertainty budget for line-intensity mapping with CONCERTO.

We stress that these values quantify relative variations of the flat-field pattern between observing conditions, rather than absolute calibration uncertainties. The high correlation coefficients indicate that the spatial response pattern is largely preserved. Such residual variations are small compared to the dominant atmospheric and foreground contributions in line-intensity mapping and are further mitigated by combining many detectors and scans in the map-making process.
Day-to-day comparisons further confirm this stability, with rms differences below 0.05 and correlation coefficients above 0.96.

After the initial detector selection (see Sect.\,\ref{sect:data_red}), about 1300 LEKIDs were retained for further analysis. We then calculated the flat-field and applied an additional validity criterion to ensure robust calibration: for each detector, the rms difference relative to reference flat-fields had to remain below 0.07. This second cut excluded a further 40–50 LEKIDs that showed inconsistent or unstable responses across slices. For the detectors that passed both stages of selection, the flat-field values cluster tightly around unity, with ranges of 0.3–1.6 (standard deviation 0.17) for LF and 0.4–1.6 (standard deviation 0.21) for HF. These distributions reflect the intrinsic spread of detector responses while confirming the overall stability of the calibration. Some of the detectors excluded in the second stage can be seen in Fig.\,\ref{fig:flat-fields}, which illustrates the flat-field distribution before applying this final cut.

\label{subsec:inv2spec}

\begin{figure*}[ht]
    \centering
    \resizebox{0.95\textwidth}{!}{
        \begin{tikzpicture}[
            node distance=0.6cm,
            >=stealth,
            block/.style={
                rectangle, 
                rounded corners, 
                draw=black, 
                thick,
                align=center,
                minimum height=1cm,
                minimum width=3.2cm
            },
        input/.style={trapezium,trapezium stretches=true,draw,thick,minimum width=2.5cm,minimum height=1cm,trapezium left angle=70,trapezium right angle=110,align=center},
        data/.style={trapezium,trapezium stretches=true,draw,thick,fill=gray!10,minimum width=2.5cm,minimum height=1cm,trapezium left angle=110,trapezium right angle=70,align=center},
        process/.style={rectangle, rounded corners, draw, thick, fill=gray!30, align=center,  minimum height=1cm, minimum width=2.5cm},
        database/.style = {cylinder, cylinder uses custom fill, cylinder body fill=gray!10, cylinder end fill=gray!10, shape border rotate=90, aspect=0.25, draw}
        ]
            \node[input] (interf) {Interferogram timeline data};
            \node[input, above=of interf] (zpd) {ZPD results};
            \node[input, below=of interf] (ff) {Flatfield results};
            \node[process, right=of interf] (regrid) {Regrid RA/Dec/OPD};
            \node[data, right=of regrid] (cube) {Interferogram \\ cube};
            \node[process, right=of cube] (apod) {Apodization \\ (Hanning)};
            \node[process, right=of apod] (fft) {Inverse FFT \\ (\texttt{np.fft.irfft})};
            \node[data, right=of fft] (spec) {Spectral cube};
            
            \draw[->, thick] (interf) -- (regrid);
            \draw[->, thick] (zpd) -- (regrid);
            \draw[->, thick] (ff) -- (regrid);
            \draw[->, thick] (regrid) -- (cube);
            \draw[->, thick] (cube) -- (apod);
            \draw[->, thick] (apod) -- (fft);
            \draw[->, thick] (fft) -- (spec);
        \end{tikzpicture}
    }
    \caption{Schematic illustration of the interferogram inversion pipeline. 
    Interferograms are projected onto a RA/Dec/OPD grid (with input from ZPD and flatfield) to form an interferogram cube. The interferograms are apodized, inverted via inverse FFT, along the OPD direction to be finally transformed into spectral cubes.}
    \label{fig:inversion_pipeline}
\end{figure*}

\section{Measuring the absolute brightness calibration factors from spectral cubes }
\label{sect:spectra}

The absolute brightness calibration factor is essential to translate interferometric timelines into astrophysical observables. Standard approaches rely on dedicated calibration procedures, such as planet observations, skydips, or shutter measurements. These approaches, however, require external information (e.g., a model for the planet spectrum) or additional observing time and do not always capture the conditions of the science observations themselves (e.g. when using the shutter).

We introduce here a new method to recover the absolute calibration factor directly from science data, using the differential dependence of the measured spectra on elevation.
In practice, the required airmass range is provided naturally by standard science observations, as the COSMOS field is tracked over several hours from low elevation through transit and back down, without the need for dedicated skydip measurements.
Applied to fields without strong line emission, such as COSMOS, this approach effectively transforms regular science scans into calibration datasets. Beyond their immediate application to CONCERTO, it is broadly applicable to any instrument that combines wide-field spectroscopy with variable atmospheric transmission. 

To make this explicit, once all LEKIDs are combined using the spectroscopic flat-field derived earlier, and after removing the contributions from atmospheric emission, the reference source, and stray light, the measurement (Eq.~\ref{eq:miv}) can be rewritten as:
\begin{equation}
m^\prime_{\nu}=  c^\prime_{\nu} e^{\amtn} \tcos \, ,
\label{eq:general}
\end{equation}
where $c^\prime_{\nu} = k\,\bandpass$ is the effective calibration factor directly applicable to the data, incorporating both the bandpass $\bandpass$ and the absolute calibration factor $k$. 
The goal of this section is to show how $c^\prime_{\nu}$ can be derived.

An important consequence is that the slope of the signal variation with atmospheric conditions directly traces the product $k\,\bandpass$. 
Therefore the effective calibration spectrum — and thus the instrumental bandpass shape up to an overall normalisation — can be determined directly from science observations. 
This behaviour is demonstrated explicitly using synthetic spectra in Appendix~\ref{app:synthetic_validation}.

\subsection{The airmass method \label{sect:airmass}}
The airmass method takes advantage of the dependence of atmospheric transmission on airmass to determine both the absolute calibration factors and the instrumental bandpass. Differentiating Eq.~\ref{eq:epsmethod}, where spectroscopic flat-field is applied, with respect to $AM$ yields the slope $\slope$:
\begin{equation}\label{eq:slope}
\dv{m_{\nu}}{AM}= \slope = k \bandpass \tau_\nu e^{\amtn} \tatm \,.
\end{equation}

The slope $\slope$ can be measured directly at a given frequency from different airmasses. Physically, it quantifies how the atmospheric emission changes with airmass, thus providing a lever to extract the calibration factors.

Using the slope definition (Eq.~\ref{eq:slope}), the effective calibration factor can be related to $\slope$ as 
\begin{equation}\label{eq:calfact}
c^\prime_{\nu} = \frac{\slope}{\tatm \tau_\nu\ e^{\amtn}}\,.
\end{equation}

Thus measuring the slope $\slope$ relative to the given $\tatm$, will allow to derive the effective calibration factor.

\subsection{The emissivity method \label{sect:emissivity}}
Alternatively, we can use a slightly different method which consists of deriving Eq.\,\ref{eq:epsmethod} with respect to $\epsilon_\nu$. We obtain:
\begin{equation}\label{eq:slopeeps}
    \slopeeps = \dv{m_{\nu}}{\epsilon_\nu}=k \bandpass \tatm \,.
\end{equation}
\noindent and the effective calibration factor can then be written as :
\begin{equation}\label{eq:calfacte}
c^\prime_{\nu} = \frac{\slopeeps}{\tatm} \,.
\end{equation}

Furthermore, this method allows us to rewrite Eq.\,\ref{eq:epsmethod} as a linear relation:
\begin{equation}\label{eq:epsmethod_simpl}
    m_{\nu} = \slopeeps \epsilon_\nu + m_{\nu}^0
\end{equation}
with 
\begin{equation}
    m_{\nu}^0 = -k \bandpass (\tref+\tstr).
\end{equation}
From this expression it follows that:
\begin{equation}\label{eq:epsmethod_tr}
    \ttot \equiv \tref + \tstr = - \frac{m_{\nu}^0}{\slopeeps} \tatm
\end{equation}

Compared to the airmass method, the emissivity method re-parameterizes the measurement in terms of atmospheric emissivity $\epsilon_\nu$ instead of airmass. This substitution often leads to a more direct linear relation between the signal and atmospheric conditions, simplifying the extraction of the calibration factors. While both approaches are formally consistent, the emissivity method can offer practical advantages in cases where opacity and emissivity are more robustly constrained than the explicit airmass dependence. 

The effective atmospheric brightness temperature $T^{\rm atm}$ and opacity $\tau_\nu$ are obtained from the ATM model using the PWV measured continuously by the APEX radiometer \citep{sarazin2013}. During the scans used here, PWV typically varies by less than 0.2\,mm, resulting in only small changes in the derived atmospheric spectra.

Uncertainties in $T^{\rm atm}$ mainly affect the overall absolute calibration scale, since both calibration methods depend on $1/T^{\rm atm}$. The dominant source of uncertainty in $T^{\rm atm}$ arises from the determination of the water vapour column, while the intrinsic accuracy of the ATM model itself is at the level of $\sim1$\% (Pardo, priv.\ comm.).

\begin{figure*}[ht]
\centering
\includegraphics[width=.45\textwidth]{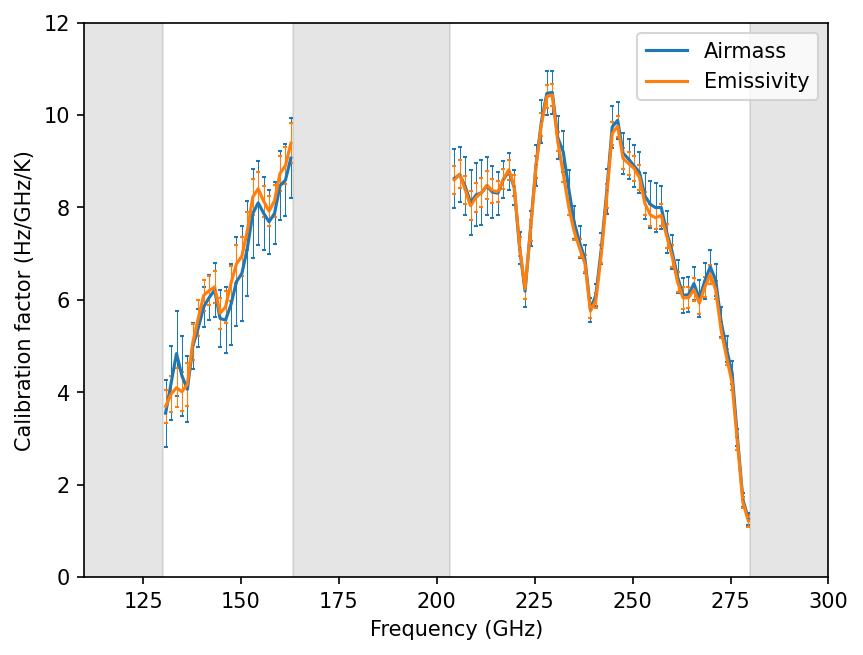}
\includegraphics[width=.45\textwidth]{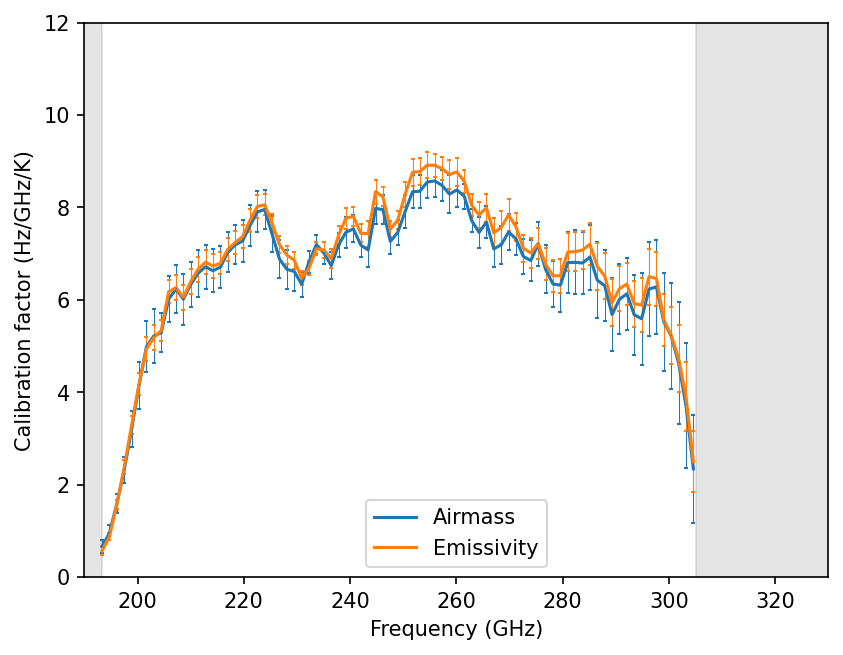}
\caption{Comparison of effective absolute spectral brightness calibration factors ($k\,\bandpass$) derived using the airmass method (Eq.\,\ref{eq:calfact}) and the emissivity method (Eq.\,\ref{eq:calfacte}) for the LF (left) and HF (right) arrays. Deviations near strong atmospheric lines are expected (masked by shaded areas), and these regions are excluded from the scientific analysis. Error bars indicate the $1\sigma$ uncertainties on the calibration factors obtained by propagating the uncertainties of the fitted slopes in the airmass and emissivity relations. The two methods yield very similar calibration spectra, agreeing within about 5\% across the usable frequency range (excluding the shaded regions where the fits are less reliable).
}
\label{fig:cf}
\end{figure*}

For illustration, a conservative variation of $\pm10$\,K around the typical value $T^{\rm atm}\simeq258$\,K would translate into a $\sim4$\% change in the derived calibration factors. Such a shift affects both the airmass and emissivity methods identically and therefore does not impact their relative comparison (Fig.~\ref{fig:cf}). The resulting absolute calibration uncertainty is consistent with the independent photometric calibration reported in \citet{hu2024}.

\subsection{Practical implementation}

For our analysis, we selected COSMOS scans away from transit (to maximise the range in airmass) and divided them into slices of 50 interferogram blocks (see Sect\,\ref{sect:obs}).
Each slice was processed into an interferogram cube by combining all valid detectors using the spectroscopic diffuse flatfield and ZPD discussed earlier. The procedure is illustrated schematically in Fig.~\ref{fig:inversion_pipeline}. Starting from the interferogram timeline data, we correct for the ZPD and apply detector flat-fielding, both obtained using the forward model. The data are then rebinned in a common optical path difference (OPD) and projected onto a regular sky grid in right ascension and declination, producing interferogram cubes. These cubes are apodized along the OPD axis, typically with a Hanning window, and subsequently inverted to spectrum space via an inverse Fourier transform. 
The result is a three-dimensional spectral cube, expressed in detector units as a function of sky position and electromagnetic frequency. 

Each 25-minute scan is divided into approximately 60 slices of 50 interferogram blocks, corresponding to time intervals of about 25–30\,s per slice. The slope determination therefore relies on a sequence of short, quasi-instantaneous measurements rather than on a single long integration. During a typical scan, PWV variations are modest compared to the elevation-driven signal change; for example, in a representative COSMOS scan the median PWV is 2.96\,mm with a standard deviation of 0.05\,mm (minimum 2.87\,mm, maximum 3.11\,mm). As a result, the dominant variation of the signal arises from the systematic change in elevation, while residual atmospheric fluctuations mainly contribute to the scatter of the fit.

To validate the method, we first applied it to synthetic spectra replicating typical observing conditions (see Appendix~\ref{app:synthetic_validation}). The procedure successfully recovered instrument properties except near strong atmospheric lines. In these regions the assumption of linear spectral variation with airmass is no longer valid, and convolution of the strong lines with the instrumental sampling introduces spurious features that prevent reliable recovery.

\subsection{Absolute spectral brightness calibration}
\label{sect_abscal}

The absolute calibration of CONCERTO spectroscopic data for point sources will be derived from Mars observations  (Beelen et al.\ in prep.) by comparing dedicated calibration scans with the model of \citet{Lellouch}.

In the previous subsections, we presented two alternative approaches for determining absolute calibration factors that are directly applicable to diffuse emission: the airmass method (Eq.\,\ref{eq:calfact}) and the emissivity method (Eq.\,\ref{eq:calfacte}), which do not require prior knowledge of the precise instrument bandpasses.
In Fig.\,\ref{fig:cf}, we compare the effective calibration factors obtained with these two approaches for both LF and HF arrays. 
The figure demonstrates that both methods provide very consistent results, strengthening confidence in their use for extended sources.

In \citet{hu2024}, the absolute photometric calibration factors were derived by observing Uranus and comparing the measured flux to a well-established planetary model. This yielded calibration factors of 25.6$\pm$0.9 Hz (Jy beam$^{-1}$)$^{-1}$ and  19.5$\pm$0.6 Hz (Jy beam$^{-1}$)$^{-1}$ for the LF and HF channels, respectively (see their Fig.~12). We can do a rough consistency check between these calibration factors (point source, photometric mode) and those derived here (extended emission, spectroscopic mode).
We first converted calibration factors from \citet{hu2024} into Hz/GHz/K using the effective beam solid angles $\Omega_{\rm eff}$ reported for 2022 in their Table 3, together with the conversion between MJy/sr and K provided in their Appendix C. The resulting values are 4.31\,Hz/GHz/K and 4.41\,Hz/GHz/K for LF and HF, respectively. 
Then these values have to be corrected for the interferometric efficiency, which quantifies the interferometric signal lost in the optics due to optical misalignment and chromatic effects in the optical components. As a first approximation, preliminary laboratory tests across the array indicate an interferometric efficiency of about 0.7.
This gives calibration factors of $\sim$6.2 Hz/GHz/K for both LF and HF, which compare well with our determinations (Fig.~\ref{fig:cf}).

\subsection{Independent estimate of the reference and stray-light temperatures}

The emissivity method also provides an estimate of the combined temperature of the internal reference source and stray-light contribution (see Eq.\,\ref{eq:epsmethod_tr}). This constitutes an independent, on-sky measurement that can be compared directly with the values adopted for the model (Sects\,\ref{subsec:refsource} and \ref{subsec:slc}), thereby serving as a verification of the accuracy of the model. 
These estimates rely on the intercept of the emissivity fit (Eq.~\ref{eq:epsmethod_simpl}), which corresponds to the combined reference-source and stray-light contribution. This offset is more sensitive to residual baseline fluctuations and modelling uncertainties than the slope, and its determination is therefore comparatively noisy.
We derive averaged values (from 220 to 270\,GHz) of 71.3$\pm$3.4\,K for LF and HF. This is slightly below but within the errors of the theoretical model, 74.2$\pm$0.7\,K, validating both the modelling inputs and the emissivity method.

\section{Conclusion}
\label{sect:cl}
We have developed and applied an observation-driven forward-modelling framework to characterise and calibrate the CONCERTO spectrometer on APEX. By modelling the dominant instrumental and atmospheric contributions directly at the interferogram level, we establish a coherent calibration chain linking interferogram reconstruction to spectrum-space calibration. This approach enables robust determination of the zero path difference (ZPD), the relative detector responses, and the absolute spectral calibration factors directly from science observations.

Quantitatively, we find that the ZPD exhibits systematic, elevation-dependent shifts of up to $\sim$0.5\,mm across the array, attributable to small optical misalignments and structural deformations. Correcting these on a per-interferogram basis reduces residual errors to below 0.02\,mm, avoiding amplitude losses in the reconstructed spectra. The flat-field derived from the response shows systematic variations with detector position but is highly stable with time and observing conditions: comparisons across elevation and PWV regimes yield rms differences of 0.04 (LF) and 0.042 (HF), mean absolute differences of 0.025 (LF) and 0.026 (HF), and correlation coefficients of 0.97 (LF) 0.98 (HF). These results demonstrate that the flat-field is essentially invariant over typical runs.

We further established two complementary absolute calibration routes that exploit atmospheric variation: the airmass method and the emissivity method. Both yield consistent effective calibration factors, agreeing within 5\% across LF and HF bands.  
This validates the robustness of the forward-model approach and enables a direct conversion from detector units to astrophysical brightness units, without requiring detailed prior knowledge of the instrumental bandpasses. This is a real asset for a line intensity mapping experiment.

Together, these results establish a coherent end-to-end calibration scheme for CONCERTO, linking interferogram correction (ZPD and flat-field) to the construction of calibrated spectral cubes with quantified systematics.

Future refinements will focus on 
(i) a formal uncertainty budget combining atmospheric and instrumental residuals,
(ii) systematic comparison of the two calibration methods across the full range of atmospheric conditions sampled by the observations, 
(iii) improved modelling of elevation-dependent stray light, and
(iv) a deeper characterisation of OPD deformations to better capture elevation-dependent effects.

Our methodology provides a solid foundation for the development and exploitation of future instruments of the same type, establishing a calibration framework that can inform the design and analysis of next-generation LEKID-based spectrometers.

\begin{acknowledgements}
We thank the anonymous referee for the careful reading and insightful comments, which have significantly improved the clarity and presentation of the article.
We would also like to acknowledge the many technicians and engineers who contributed to the development of the CONCERTO experimental setup and its successful operation: Maurice Grollier, Olivier Exshaw, Anne Gerardin, Gilles Pont, Guillaume Donnier-Valentin, Philippe Jeantet, Mathilde Heigeas, Christophe Vescovi, Marc Marton, Christophe Hoarau, Jean-Paul Leggeri, Julien Marpaud, Samuel Roni, Damien Tourres, Sebastien Roudier, and Guillaume Bres. 
We acknowledge the crucial contributions of the whole Cryogenics and Electronics groups at Institut Néel and LPSC. We acknowledge the contributions of Hamdi Mani, Chris Groppi, and Philip Mauskopf (from the School of Earth and Space Exploration and the Department of Physics, Arizona State University) to cold electronics. The KID arrays of CONCERTO have been produced at the PTA Grenoble microfabrication facility. We warmly thank the support from the APEX staff for their support during observations and in CONCERTO pre-installations and design. The flexible pipes, in particular, have been routed under the competent coordination of Jorge Santana and Marcelo Navarro.
We are grateful to our administrative staff in Grenoble and Marseille, in particular Patricia Poirier, Mathilde Berard, Lilia Todorov, and Valérie Favre, and the Protisvalor team. We acknowledge the crucial help of the Institut Néel and MCBT Heads (Etienne Bustarret, Klaus Hasselbach, Thierry Fournier, Laurence Magaud) during the COVID-19 restriction period.
Based on observations with the APEX telescope under programme ID 108.21X2.001. APEX is a collaboration between the Max-Planck-Institut fuer Radioastronomie, the European Southern Observatory, and the Onsala Observatory.
This work has been supported by the LabEx FOCUS ANR-11-LABX-0013, the European Research Council (ERC) under the European Union's Horizon 2020 research and innovation programme (project CONCERTO, grant agreement No 788212), and the Excellence Initiative of Aix-Marseille University-A*Midex, a French ``Investissements d'Avenir'' programme.
This work has also been supported by the GIS KIDs.
MA is supported by FONDECYT grant number 1252054, and gratefully acknowledges support from ANID Basal Project FB210003 and ANID MILENIO NCN2024\_112.
This research made use of {\tt Astropy} (\url{http://www.astropy.org}), a community-developed core Python package for Astronomy \citep{astropy:2013, astropy:2018}. We also use  {\tt Matplotlib} (\url{https://matplotlib.org}, \citep{Hunter:2007}), {\tt NumPy} (\url{https://numpy.org}, \citealp{harris2020array}) and {\tt SciPy} (\url{http://www.scipy.org}, \citealp{2020SciPy-NMeth}).
\end{acknowledgements}

\bibliographystyle{bibtex/aa}
\bibliography{bibliography.bib}

@ARTICLE{sarazin2013,
       author = {{Sarazin}, M. and {Kerber}, F. and {De Breuck}, C.},
        title = "{Precipitable Water Vapour at the ESO Observatories: The Skill of the Forecasts}",
      journal = {The Messenger},
         year = 2013,
        month = jun,
       volume = {152},
        pages = {17-21},
       adsurl = {https://ui.adsabs.harvard.edu/abs/2013Msngr.152...17S}
}

@ARTICLE{pardo2025,
       author = {{Pardo}, J.~R. and {De Breuck}, C. and {Muders}, D. and {Gonz{\'a}lez}, J. and {P{\'e}rez-Beaupuits}, J.~P. and {Cernicharo}, J. and {Prigent}, C. and {Serabyn}, E. and {Montenegro-Montes}, F.~M. and {Mroczkowski}, T. and {Phillips}, N. and {Villard}, E.},
        title = "{Validation of millimetre and sub-millimetre atmospheric collision-induced absorption at Chajnantor}",
      journal = {\aap},
         year = 2025,
        month = jan,
       volume = {693},
          eid = {A148},
        pages = {A148},
          doi = {10.1051/0004-6361/202452159},
archivePrefix = {arXiv},
       eprint = {2411.03134},
 primaryClass = {astro-ph.IM},
       adsurl = {https://ui.adsabs.harvard.edu/abs/2025A&A...693A.148P}
}

@misc{Lellouch,
       author = {{Lellouch}, E. and {Amri}, H.},
       year = 2008,
       title = "{Mars Brightness Model}",
       howpublished = {\url{http://www.lesia.obspm.fr/perso/emmanuel-lellouch/mars/}},
       note = {Accessed: 2023-05-30}
}

@INPROCEEDINGS{fasano2024,
       author = {{Fasano}, Alessandro and {Ade}, Peter and {Aravena}, Manuel and {Barria}, Emilio and {Beelen}, Alexandre and {Beno{\^\i}t}, Alain and {B{\'e}thermin}, Matthieu and {Bounmy}, Julien and {Bourrion}, Olivier and {Bres}, Guillaume and et al.},
        title = "{CONCERTO: instrument model of Fourier transform spectroscopy, white-noise components}",
    booktitle = {Millimeter, Submillimeter, and Far-Infrared Detectors and Instrumentation for Astronomy XII},
         year = 2024,
       editor = {{Zmuidzinas}, Jonas and {Gao}, Jian-Rong},
       series = {Society of Photo-Optical Instrumentation Engineers (SPIE) Conference Series},
       volume = {13102},
        month = aug,
          eid = {131020O},
        pages = {131020O},
          doi = {10.1117/12.3019890},
       adsurl = {https://ui.adsabs.harvard.edu/abs/2024SPIE13102E..0OF}
}

@INPROCEEDINGS{fasano2022,
       author = {{Fasano}, Alessandro and {Beelen}, Alexandre and {Beno{\^\i}t}, Alain and {Lundgren}, Andreas and {Ade}, Peter and {Aravena}, Manuel and {Barria}, Emilio and {B{\'e}thermin}, Matthieu and {Bounmy}, Julien and {Bourrion}, Olivier and {Bres}, Guillaume and {Calvo}, Martino and {Catalano}, Andrea and {D{\'e}sert}, Fran{\c{c}}ois-Xavier and {De Breuck}, Carlos and {Dur{\'a}n}, Carlos and {Fenouillet}, Thomas and {Garcia}, Jose and {Garde}, Gregory and {Goupy}, Johannes and {Groppi}, Christopher and {Hoarau}, Christophe and {Hu}, Wenkai and {Lagache}, Guilaine and {Lambert}, Jean-Charles and {Leggeri}, Jean-Paul and {Levy-Bertrand}, Florence and {Mac{\'\i}as-P{\'e}rez}, Juan and {Mani}, Hamdi and {Marpaud}, Julien and {Mauskopf}, Philip and {Monfardini}, Alessandro and {Pisano}, Giampaolo and {Ponthieu}, Nicolas and {Prieur}, Leo and {Roni}, Samuel and {Roudier}, Sebastien and {Tourres}, Damien and {Tucker}, Carol and {Van Cuyck}, Mathilde},
        title = "{CONCERTO: a breakthrough in wide field-of-view spectroscopy at millimeter wavelengths}",
    booktitle = {Millimeter, Submillimeter, and Far-Infrared Detectors and Instrumentation for Astronomy XI},
         year = 2022,
       editor = {{Zmuidzinas}, Jonas and {Gao}, Jian-Rong},
       series = {Society of Photo-Optical Instrumentation Engineers (SPIE) Conference Series},
       volume = {12190},
        month = aug,
          eid = {121900Q},
        pages = {121900Q},
          doi = {10.1117/12.2629228},
archivePrefix = {arXiv},
       eprint = {2206.15146},
 primaryClass = {astro-ph.IM},
       adsurl = {https://ui.adsabs.harvard.edu/abs/2022SPIE12190E..0QF}
}

@ARTICLE{desert2025,
       author = {{D{\'e}sert}, F.-X. and {Mac{\'\i}as-P{\'e}rez}, J.~F. and {Beelen}, A. and {Beno{\^\i}t}, A. and {B{\'e}thermin}, M. and {Bounmy}, J. and {Bourrion}, O. and {Calvo}, M. and {Catalano}, A. and {De Breuck}, C. and {Dubois}, C. and {Dur{\'a}n}, C.~A. and {Fasano}, A. and {Goupy}, J. and {Hu}, W. and {Ibar}, E. and {Lagache}, G. and {Lundgren}, A. and {Monfardini}, A. and {Ponthieu}, N. and {Quinatoa}, D. and {Van Cuyck}, M. and {Adam}, R. and {Ade}, P. and {Ajeddig}, H. and {Amarantidis}, S. and {Andr{\'e}}, P. and {Aussel}, H. and {Berta}, S. and {Bongiovanni}, A. and {Ch{\'e}rouvrier}, D. and {De Petris}, M. and {Doyle}, S. and {Driessen}, E.~F.~C. and {Ejlali}, G. and {Ferragamo}, A. and {Gomez}, A. and {Hanser}, C. and {Katsioli}, S. and {K{\'e}ruzor{\'e}}, F. and {Kramer}, C. and {Ladjelate}, B. and {Leclercq}, S. and {Lestrade}, J.-F. and {Madden}, S.~C. and {Maury}, A. and {Mayet}, F. and {Moyer-Anin}, A. and {Mu{\~n}oz-Echeverr{\'\i}a}, M. and {Myserlis}, I. and {Paliwal}, A. and {Perotto}, L. and {Pisano}, G. and {Rev{\'e}ret}, V. and {Rigby}, A.~J. and {Ritacco}, A. and {Roussel}, H. and {Ruppin}, F. and {S{\'a}nchez-Portal}, M. and {Savorgnano}, S. and {Sievers}, A. and {Tucker}, C. and {Zylka}, R.},
        title = "{Continuum, CO, and water vapour maps of the Orion Nebula: First millimetre spectral imaging with CONCERTO}",
      journal = {\aap},
         year = 2025,
        month = sep,
       volume = {701},
          eid = {A210},
        pages = {A210},
          doi = {10.1051/0004-6361/202555320},
archivePrefix = {arXiv},
       eprint = {2504.20487},
 primaryClass = {astro-ph.GA},
       adsurl = {https://ui.adsabs.harvard.edu/abs/2025A&A...701A.210D}
}

@ARTICLE{2025JCAP...04..020K,
       author = {{Kogut}, Alan and {Aghanim}, Nabila and {Chluba}, Jens and others},
        title = "{The Primordial Inflation Explorer (PIXIE): mission design and science goals}",
      journal = {\jcap},
         year = 2025,
        month = apr,
       volume = {2025},
       number = {4},
          eid = {020},
        pages = {020},
          doi = {10.1088/1475-7516/2025/04/020},
archivePrefix = {arXiv},
       eprint = {2405.20403},
 primaryClass = {astro-ph.CO},
       adsurl = {https://ui.adsabs.harvard.edu/abs/2025JCAP...04..020K}
}

@ARTICLE{hu2024,
       author = {{Hu}, W. and {Beelen}, A. and {Lagache}, G. and {Fasano}, A. and {Lundgren}, A. and {Ade}, P. and {Aravena}, M. and {Barria}, E. and {Benoit}, A. and {B{\'e}thermin}, M. and {Bounmy}, J. and {Bourrion}, O. and {Bres}, G. and {De Breuck}, C. and {Calvo}, M. and {Catalano}, A. and {D{\'e}sert}, F. -X. and {Dubois}, C. and {Dur{\'a}n}, C.~A. and {Fenouillet}, T. and {Garcia}, J. and {Garde}, G. and {Goupy}, J. and {Hoarau}, C. and {Lambert}, J. -C. and {Lellouch}, E. and {Levy-Bertrand}, F. and {Macias-Perez}, J. and {Marpaud}, J. and {Monfardini}, A. and {Pisano}, G. and {Ponthieu}, N. and {Prieur}, L. and {Quinatoa}, D. and {Roni}, S. and {Roudier}, S. and {Tourres}, D. and {Tucker}, C. and {Van Cuyck}, M.},
        title = "{CONCERTO at APEX On-sky performance in continuum}",
      journal = {\aap},
         year = 2024,
        month = sep,
       volume = {689},
          eid = {A20},
        pages = {A20},
          doi = {10.1051/0004-6361/202449260},
archivePrefix = {arXiv},
       eprint = {2406.15572},
 primaryClass = {astro-ph.IM},
       adsurl = {https://ui.adsabs.harvard.edu/abs/2024A&A...689A..20H}
}

@ARTICLE{2024PASP..136k4505M,
       author = {{Mac{\'\i}as-P{\'e}rez}, J.~F. and {Fern{\'a}ndez-Torreiro}, M. and {Catalano}, A. and {Fasano}, A. and others},
        title = "{KISS: Instrument Description and Performance}",
      journal = {\pasp},
         year = 2024,
        month = nov,
       volume = {136},
       number = {11},
          eid = {114505},
        pages = {114505},
          doi = {10.1088/1538-3873/ad8189},
archivePrefix = {arXiv},
       eprint = {2409.20272},
 primaryClass = {astro-ph.IM},
       adsurl = {https://ui.adsabs.harvard.edu/abs/2024PASP..136k4505M}
}

@ARTICLE{2021NatAs...5..970M,
       author = {{Monfardini}, Alessandro and {Lagache}, Guilaine},
        title = "{A magnum opus on the Chajnantor plateau}",
      journal = {Nature Astronomy},
         year = 2021,
        month = sep,
       volume = {5},
        pages = {970-970},
          doi = {10.1038/s41550-021-01482-1},
       adsurl = {https://ui.adsabs.harvard.edu/abs/2021NatAs...5..970M}
}

@INPROCEEDINGS{2024SPIE13102E,
       author = {{Maffei}, B. and {Aghanim}, N. and {Aumont}, J. and others},
        title = "{BISOU: a balloon pathfinder for CMB spectral distortions studies}",
    booktitle = {Society of Photo-Optical Instrumentation Engineers (SPIE) Conference Series},
         year = 2024,
       editor = {{Zmuidzinas}, Jonas and {Gao}, Jian-Rong},
       series = {Society of Photo-Optical Instrumentation Engineers (SPIE) Conference Series},
       volume = {13102},
        month = aug,
          eid = {131020N},
        pages = {131020N},
          doi = {10.1117/12.3018371},
       adsurl = {https://ui.adsabs.harvard.edu/abs/2024SPIE13102E..0NM}
}

@ARTICLE{2022JInst..17P0047B,
       author = {{Bourrion}, O. and {Hoarau}, C. and {Bounmy}, J. and others},
        title = "{CONCERTO: readout and control electronics}",
      journal = {Journal of Instrumentation},
         year = 2022,
        month = oct,
       volume = {17},
       number = {10},
          eid = {P10047},
        pages = {P10047},
          doi = {10.1088/1748-0221/17/10/P10047},
archivePrefix = {arXiv},
       eprint = {2208.07629},
 primaryClass = {astro-ph.IM},
       adsurl = {https://ui.adsabs.harvard.edu/abs/2022JInst..17P0047B}
}

@ARTICLE{2022JInst..17P8037B,
       author = {{Bounmy}, Julien and {Hoarau}, Christophe and {Mac{\'\i}as-P{\'e}rez}, Juan-Francisco and others},
        title = "{CONCERTO: Digital processing for finding and tuning LEKIDs}",
      journal = {Journal of Instrumentation},
         year = 2022,
        month = aug,
       volume = {17},
       number = {8},
          eid = {P08037},
        pages = {P08037},
          doi = {10.1088/1748-0221/17/08/P08037},
archivePrefix = {arXiv},
       eprint = {2206.11554},
 primaryClass = {astro-ph.IM},
       adsurl = {https://ui.adsabs.harvard.edu/abs/2022JInst..17P8037B}
}

@ARTICLE{fasano2021,
       author = {{Fasano}, A. and {Mac{\'\i}as-P{\'e}rez}, J.~F. and {Benoit} and others},
        title = "{Accurate sky signal reconstruction for ground-based spectroscopy with kinetic inductance detectors}",
      journal = {\aap},
         year = 2021,
        month = dec,
       volume = {656},
          eid = {A116},
        pages = {A116},
          doi = {10.1051/0004-6361/202141419},
archivePrefix = {arXiv},
       eprint = {2109.03145},
 primaryClass = {astro-ph.IM},
       adsurl = {https://ui.adsabs.harvard.edu/abs/2021A&A...656A.116F}
}

@ARTICLE{2020SciPy-NMeth,
  author  = {Virtanen, Pauli and Gommers, Ralf and Oliphant, Travis E. and
            others},
  title   = {{{SciPy} 1.0: Fundamental Algorithms for Scientific
            Computing in Python}},
  journal = {Nature Methods},
  year    = {2020},
  volume  = {17},
  pages   = {261--272},
  adsurl  = {https://rdcu.be/b08Wh},
  doi     = {10.1038/s41592-019-0686-2},
}

@Article{Hunter:2007,
  Author    = {Hunter, J. D.},
  Title     = {Matplotlib: A 2D graphics environment},
  Journal   = {Computing in Science \& Engineering},
  Volume    = {9},
  Number    = {3},
  Pages     = {90--95},
  publisher = {IEEE COMPUTER SOC},
  doi       = {10.1109/MCSE.2007.55},
  year      = 2007
}

@Article{ harris2020array,
     title = {Array programming with {NumPy}},
     author = {Charles R. Harris and K. Jarrod Millman and St{\'{e}}fan J.
                     van der Walt and others},
     year          = {2020},
     month         = sep,
     journal       = {Nature},
     volume        = {585},
     number        = {7825},
     pages         = {357--362},
     doi           = {10.1038/s41586-020-2649-2},
     publisher     = {Springer Science and Business Media {LLC}},
     url           = {https://doi.org/10.1038/s41586-020-2649-2}
}

@article{astropy:2013,
    Adsurl = {http://adsabs.harvard.edu/abs/2013A%26A...558A..33A},
    Archiveprefix = {arXiv},
    Author = {{Astropy Collaboration} and others},
    Doi = {10.1051/0004-6361/201322068},
    Eid = {A33},
    Eprint = {1307.6212},
    Journal = {\aap},
    Month = oct,
    Pages = {A33},
    Primaryclass = {astro-ph.IM},
    Title = {{Astropy: A community Python package for astronomy}},
    Volume = 558,
    Year = 2013
}

@ARTICLE{astropy:2018,
       author = {{Astropy Collaboration} and others},
        title = "{The Astropy Project: Building an Open-science Project and Status of the v2.0 Core Package}",
      journal = {\aj},
         year = 2018,
        month = sep,
       volume = {156},
       number = {3},
          eid = {123},
        pages = {123},
          doi = {10.3847/1538-3881/aabc4f},
archivePrefix = {arXiv},
       eprint = {1801.02634},
 primaryClass = {astro-ph.IM},
       adsurl = {https://ui.adsabs.harvard.edu/abs/2018AJ....156..123A}
}

@INPROCEEDINGS{2010SPIE.7741E..0MD,
       author = {{Doyle}, Simon and {Mauskopf}, Philip and {Zhang}, Jin and {Monfardini}, Alessandro and {Swenson}, Loren and {Baselmans}, Jochem J.~A. and {Yates}, Stephen J.~C. and {Roesch}, Markus},
        title = "{A review of the lumped element kinetic inductance detector}",
    booktitle = {Millimeter, Submillimeter, and Far-Infrared Detectors and Instrumentation for Astronomy V},
         year = 2010,
       editor = {{Holland}, Wayne S. and {Zmuidzinas}, Jonas},
       series = {Society of Photo-Optical Instrumentation Engineers (SPIE) Conference Series},
       volume = {7741},
        month = jul,
          eid = {77410M},
        pages = {77410M},
          doi = {10.1117/12.857341},
       adsurl = {https://ui.adsabs.harvard.edu/abs/2010SPIE.7741E..0MD}
}

@ARTICLE{adam2018,
       author = {{Adam}, R. and {Adane}, A. and {Ade}, P.~A.~R. and others},
        title = "{The NIKA2 large-field-of-view millimetre continuum camera for the 30 m IRAM telescope}",
      journal = {\aap},
         year = 2018,
        month = jan,
       volume = {609},
          eid = {A115},
        pages = {A115},
          doi = {10.1051/0004-6361/201731503},
archivePrefix = {arXiv},
       eprint = {1707.00908},
 primaryClass = {astro-ph.IM},
       adsurl = {https://ui.adsabs.harvard.edu/abs/2018A&A...609A.115A}
}

@ARTICLE{2020A&A...641A.179C,
       author = {{Catalano}, A. and {Bideaud}, A. and {Bourrion}, O. and others},
        title = "{Sensitivity of LEKID for space applications between 80 GHz and 600 GHz}",
      journal = {\aap},
         year = 2020,
        month = sep,
       volume = {641},
          eid = {A179},
        pages = {A179},
          doi = {10.1051/0004-6361/202038199},
       adsurl = {https://ui.adsabs.harvard.edu/abs/2020A&A...641A.179C}
}

@ARTICLE{2020A&A...642A..60C,
       author = {{Concerto Collaboration} and {Ade}, P. and {Aravena}, M. others},
        title = "{A wide field-of-view low-resolution spectrometer at APEX: Instrument design and scientific forecast}",
      journal = {\aap},
         year = 2020,
        month = oct,
       volume = {642},
          eid = {A60},
        pages = {A60},
          doi = {10.1051/0004-6361/202038456},
       adsurl = {https://ui.adsabs.harvard.edu/abs/2020A&A...642A..60C}
}

@ARTICLE{2010A&A...518L...3G,
       author = {{Griffin}, M.~J. and {Abergel}, A. and {Abreu}, A. and {Ade}, P.~A.~R. and others},
        title = "{The Herschel-SPIRE instrument and its in-flight performance}",
      journal = {\aap},
         year = 2010,
        month = jul,
       volume = {518},
          eid = {L3},
        pages = {L3},
          doi = {10.1051/0004-6361/201014519},
archivePrefix = {arXiv},
       eprint = {1005.5123},
 primaryClass = {astro-ph.IM},
       adsurl = {https://ui.adsabs.harvard.edu/abs/2010A&A...518L...3G}
}

@ARTICLE{pardo2001,
	author = {{Pardo}, J.~R. and {Serabyn}, E. and {Cernicharo}, J.},
	title = "{Submillimeter atmospheric transmission measurements on Mauna Kea during extremely dry El Nino conditions: implications for broadband opacity contributions}",
	journal = {\jqsrt},
	year = 2001,
	month = feb,
	volume = {68},
	pages = {419-433},
	doi = {10.1016/S0022-4073(00)00034-0},
	adsurl = {https://ui.adsabs.harvard.edu/abs/2001JQSRT..68..419P}
}

@ARTICLE{fasano-ltd,
	author = {{Fasano}, A. and {Aguiar}, M. and {Benoit}, A. and others},
	title = "{The KISS Experiment}",
	journal = {Journal of Low Temperature Physics},
	year = 2020,
	month = apr,
	volume = {199},
	number = {1-2},
	pages = {529-536},
	doi = {10.1007/s10909-019-02289-1},
	archivePrefix = {arXiv},
	eprint = {1911.13148},
	primaryClass = {astro-ph.IM},
	adsurl = {https://ui.adsabs.harvard.edu/abs/2020JLTP..199..529F}
}

@ARTICLE{mpi,
	author = {{Martin}, D.~H. and {Puplett}, E.},
	title = "{Polarised interferometric spectrometry for the millimeter and submillimeter spectrum.}",
	journal = {Infrared Physics},
	year = 1970,
	volume = 10,
	pages = {105-109},
	doi = {10.1016/0020-0891(70)90006-0},
	adsurl = {http://adsabs.harvard.edu/abs/1970InfPh..10..105M}
}

@ARTICLE{FIRAS,
	author = {{Mather}, J.~C. and {Fixsen}, D.~J. and {Shafer}, R.~A. and
	{Mosier}, C. and {Wilkinson}, D.~T.},
	title = "{Calibrator Design for the COBE Far-Infrared Absolute Spectrophotometer (FIRAS)}",
	journal = {\apj},
	eprint = {astro-ph/9810373},
	year = 1999,
	month = feb,
	volume = 512,
	pages = {511-520},
	doi = {10.1086/306805},
	adsurl = {http://adsabs.harvard.edu/abs/1999ApJ...512..511M}
}

@ARTICLE{2007ApJS..172....1S,
       author = {{Scoville}, N. and {Aussel}, H. and {Brusa}, M. and {Capak}, P. and {Carollo}, C.~M. and {Elvis}, M. and {Giavalisco}, M. and {Guzzo}, L. and {Hasinger}, G. and {Impey}, C. and {Kneib}, J. -P. and {LeFevre}, O. and {Lilly}, S.~J. and {Mobasher}, B. and {Renzini}, A. and {Rich}, R.~M. and {Sanders}, D.~B. and {Schinnerer}, E. and {Schminovich}, D. and {Shopbell}, P. and {Taniguchi}, Y. and {Tyson}, N.~D.},
        title = "{The Cosmic Evolution Survey (COSMOS): Overview}",
      journal = {\apjs},
         year = 2007,
        month = sep,
       volume = {172},
       number = {1},
        pages = {1-8},
          doi = {10.1086/516585},
archivePrefix = {arXiv},
       eprint = {astro-ph/0612305},
 primaryClass = {astro-ph},
       adsurl = {https://ui.adsabs.harvard.edu/abs/2007ApJS..172....1S}
}

@ARTICLE{Monfardini2010,
	author = {{Monfardini}, A. and {Swenson}, L.~J. and {Bideaud}, A. and 
	others},
	title = "{NIKA: A millimeter-wave kinetic inductance camera}",
	journal = {\aap},
	archivePrefix = "arXiv",
	eprint = {1004.2209},
	primaryClass = "astro-ph.IM",
	year = 2010,
	month = oct,
	volume = 521,
	eid = {A29},
	pages = {A29},
	doi = {10.1051/0004-6361/201014727},
	adsurl = {http://adsabs.harvard.edu/abs/2010A%26A...521A..29M}
}

@ARTICLE{Masi_2008_OLIMPO,
	author = {{Masi}, S. and {Battistelli}, E. and {Brienza}, D. and others},
	title = "{OLIMPO}",
	journal = {\memsai},
	year = 2008,
	volume = 79,
	pages = {887},
	adsurl = {http://adsabs.harvard.edu/abs/2008MmSAI..79..887M}
}

@ARTICLE{2006A&A...454L..13G,
       author = {{G{\"u}sten}, R. and {Nyman}, L. {\r{A}}. and {Schilke}, P. and {Menten}, K. and {Cesarsky}, C. and {Booth}, R.},
        title = "{The Atacama Pathfinder EXperiment (APEX) - a new submillimeter facility for southern skies -}",
      journal = {\aap},
         year = 2006,
        month = aug,
       volume = {454},
       number = {2},
        pages = {L13-L16},
          doi = {10.1051/0004-6361:20065420},
       adsurl = {https://ui.adsabs.harvard.edu/abs/2006A&A...454L..13G}
}

\appendix
\section{Validation on synthetic spectra at fixed atmospheric conditions}
\label{app:synthetic_validation}

This appendix validates the absolute calibration methods introduced in the main text using synthetic data. 
While the results presented in the paper are derived entirely from real observations, the synthetic analysis provides a controlled test demonstrating that the airmass and emissivity methods correctly recover the calibration factors under idealised conditions.

\subsection{Construction of synthetic spectra}

Synthetic spectra were generated directly in frequency space for a fixed precipitable water vapour (PWV) of 2.9\,mm, representative of typical COSMOS observing conditions. 
For each frequency bin, the atmospheric emission and opacity were computed using the ATM model \citep{pardo2001} over a range of telescope elevations matching those sampled during the science scans.

The synthetic measurements were constructed following Eq.~\ref{eq:epsmethod} of the main text, including a representative instrumental bandpass and a fixed detector response factor. 
Only the atmospheric contribution was allowed to vary with elevation through its dependence on airmass, while the reference-source and stray-light contributions were kept constant.

Instrumental noise, bandpass uncertainties, stray-light variations, and elevation-dependent optical deformations were not included. 
The resulting dataset therefore represents an idealised case in which the assumptions underlying the airmass and emissivity calibration methods are exactly satisfied.

\subsection{Application of the absolute calibration formalism}

The synthetic spectra were analysed using the same procedures described in Sect.~\ref{sect:emissivity}. 
For each frequency bin, the detector signal was fitted as a function of atmospheric emissivity,
\begin{equation}
\epsilon_\nu = 1 - e^{-\tau_\nu \,\mathrm{AM}} ,
\end{equation}
recovering the emissivity slope defined in Eq.~\ref{eq:slopeeps}. 
This slope directly traces the effective calibration spectrum $c'_\nu = k\,\bandpass$, as defined in Sect.~\ref{sect:spectra}.

In the following subsections we first examine the linearity of the emissivity relation and then assess the recovery of the calibration factors.

\subsection{Linearity with atmospheric emissivity}

Before assessing the recovery of the calibration factors, we first verify the fundamental assumption underlying the emissivity method: that, at fixed PWV and over a limited elevation range, the detector signal varies linearly with atmospheric emissivity.

Figure~\ref{fig:synthetic_eps} shows synthetic spectra as a function of atmospheric emissivity for a fixed PWV of 2.9\,mm and an elevation range of 26–33$^\circ$. 
This relatively narrow interval corresponds to the typical elevation drift during a single 25-minute science scan. 
Each colour represents a distinct frequency channel within the LF (upper) and HF (lower) bands (blue = lowest frequency, red = highest frequency within each band), sampled at different elevations within this range. 
Identical frequency channels and colour coding are used consistently between the synthetic and real-data panels within each band.

In the synthetic case, the relation between detector signal and emissivity is strictly linear across the full elevation range. 
The fitted slopes reproduce the behaviour predicted by Eq.~\ref{eq:slopeeps}, confirming that under idealised conditions the emissivity-based formulation is mathematically exact.

\begin{figure}[t]
\centering
\includegraphics[width=.45\textwidth]{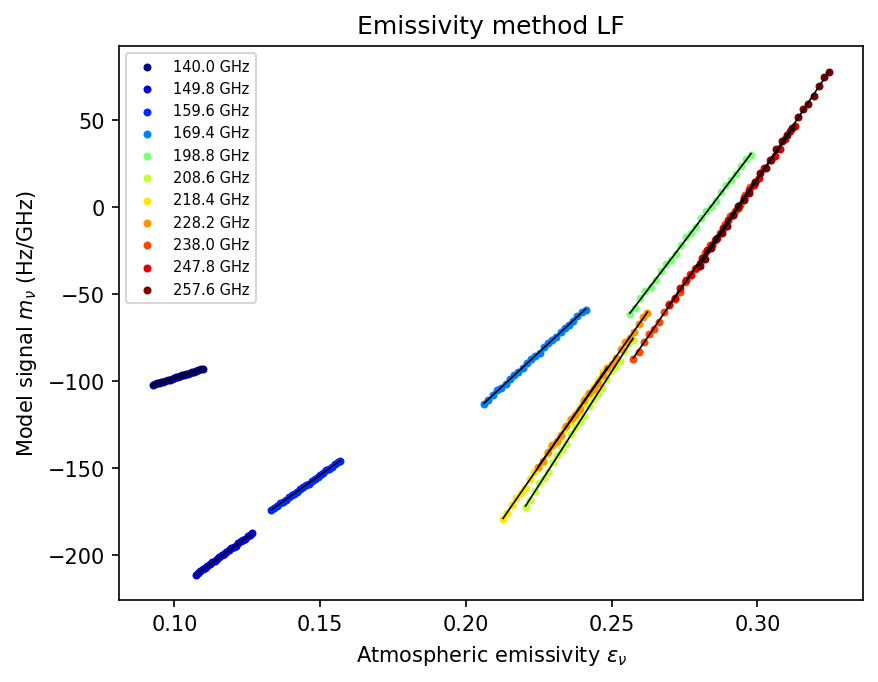}
\includegraphics[width=.45\textwidth]{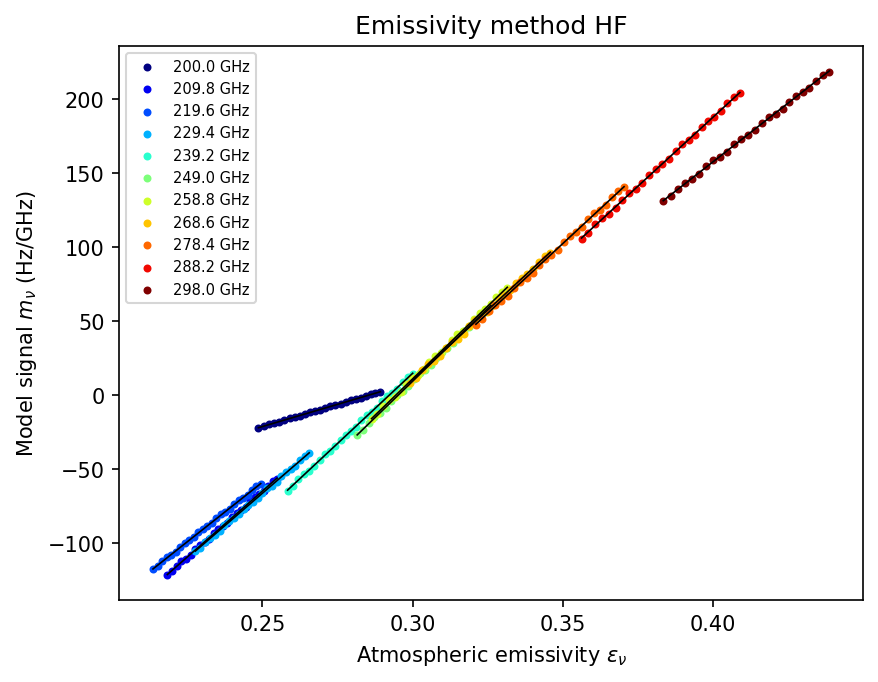}
\caption{
Synthetic spectra illustrating the linear dependence of the model signal on atmospheric emissivity under controlled conditions. 
The panels show the model signal as a function of atmospheric emissivity for representative frequency channels in the LF (upper) and HF (lower) bands for a fixed PWV of 2.9\,mm and an elevation range of 26–33$^\circ$. 
Each colour corresponds to a different frequency $\nu$, and the solid black lines show the linear fits used to derive $\mathrm{d} m_{\nu} / \mathrm{d} \epsilon_{\nu}$ (Eq.~\ref{eq:slopeeps}).
For visual comparison with the detector data, the synthetic spectra were multiplied by constant scaling factors to match the order of magnitude of the measured signal. 
Channels within $\pm$5\,GHz of the 183\,GHz water vapour line are omitted for clarity because their signals exceed the plotted dynamic range.
}
\label{fig:synthetic_eps}
\end{figure}

For comparison, Fig.~\ref{fig:real_eps} presents the same analysis applied to real CONCERTO observations at the same PWV and elevation range. 
While the data exhibit increased scatter due to instrumental noise and residual systematics, the linear trend with atmospheric emissivity remains clearly visible. 
Although the leverage in emissivity is modest (reflecting the limited elevation variation within a single scan), it is sufficient to constrain the slopes robustly.

These results validate the linear approximation underlying the emissivity method under realistic observing conditions and motivate the subsequent recovery of the effective calibration spectrum.

\begin{figure}[t]
\centering
\includegraphics[width=.44\textwidth]{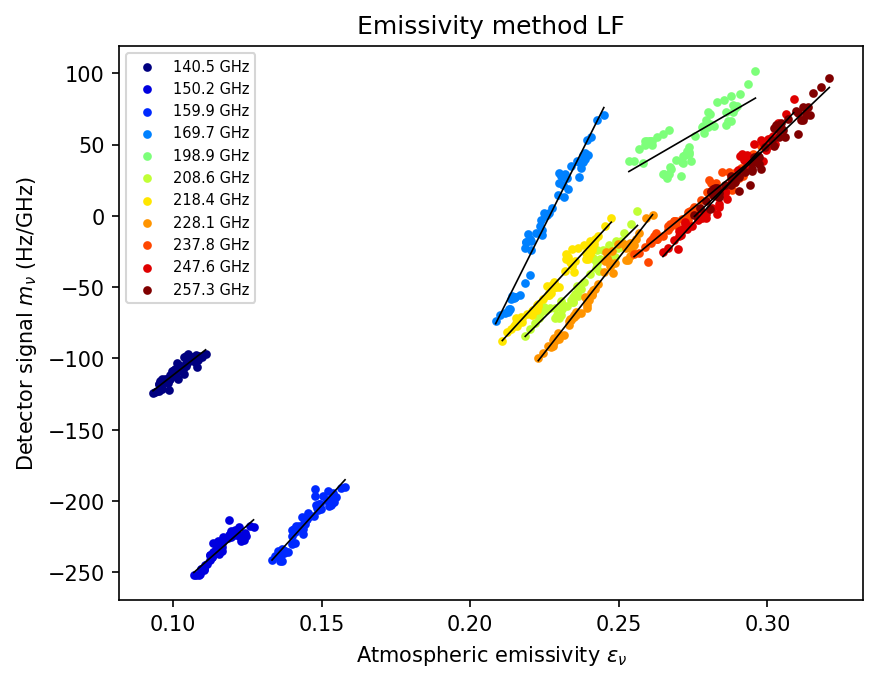}
\includegraphics[width=.44\textwidth]{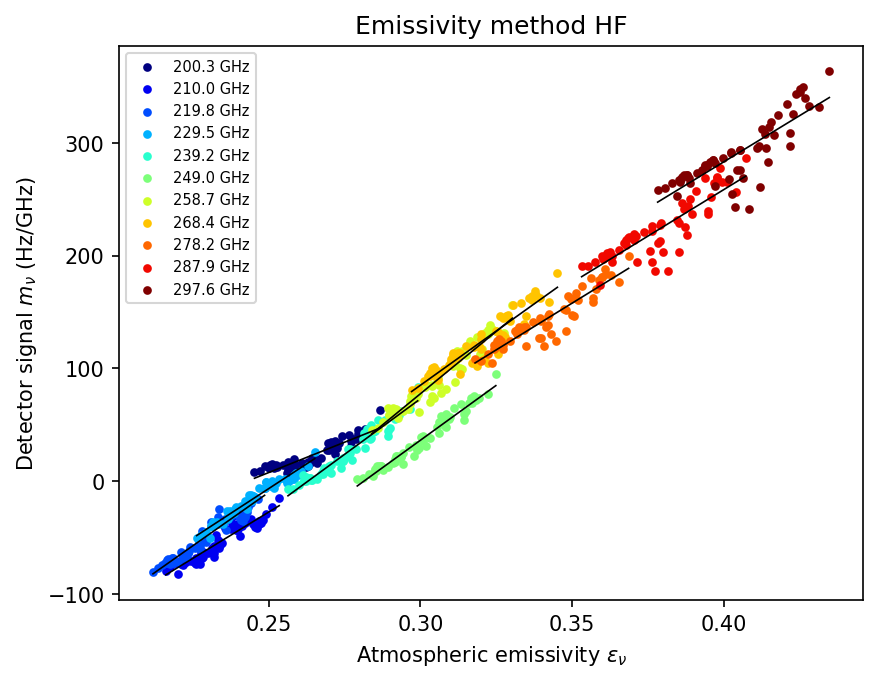}
\caption{
Measured detector signal as a function of atmospheric emissivity, for the same PWV (2.9\,mm), elevation range (26–33$^\circ$), and representative frequency channels shown in Fig.~\ref{fig:synthetic_eps}. 
Colour coding follows the same convention as in the synthetic case (blue = lowest frequency, red = highest frequency within each band). 
Linear fits are overplotted for each frequency channel, from which $\mathrm{d}m_{\nu} / \mathrm{d}\epsilon_{\nu}$ (Eq.~\ref{eq:slopeeps}) is derived.}
Compared to the synthetic case, the data exhibit increased scatter due to instrumental noise and residual systematics, while preserving the linear dependence on atmospheric emissivity.
\label{fig:real_eps}
\end{figure}

\subsection{Recovery of absolute calibration factors}

Having established the linear dependence of the detector signal on atmospheric emissivity, we now assess whether the calibration factors can be quantitatively recovered from the synthetic data.

Because the synthetic spectra are generated from a known forward model, the input effective calibration spectrum $c'_\nu = k\,\bandpass$ is known by construction. 
A representative instrumental bandpass and detector response factor were assumed to define this input spectrum.

Figure~\ref{fig:synth_cf} compares the calibration spectra recovered using the airmass and emissivity methods with the input calibration spectrum for the LF and HF bands. 
Across the usable frequency range, both methods recover the input calibration spectrum to within a few percent.

Noticeable deviations appear only in the vicinity of strong atmospheric absorption features, such as the 119\,GHz oxygen line and the 183 and 325\,GHz water vapour lines.
In these regions the assumption of linear spectral variation with airmass or emissivity breaks down, as the sharp atmospheric line profiles are convolved with the instrumental spectral response.
These frequency intervals are excluded from the scientific analysis in the main text.

We emphasise that, while the synthetic validation necessarily assumes a bandpass to construct the input spectra, the calibration factors derived from real data in Sect.~\ref{sect:spectra} do not require prior knowledge of the bandpass shape. 
In both the emissivity and airmass formulations, the recovered slopes directly trace the product $k\,\bandpass$, allowing the effective calibration spectrum — and thus the bandpass shape up to an overall normalisation — to be determined directly from science observations.

\begin{figure}
\includegraphics[width=.45\textwidth]{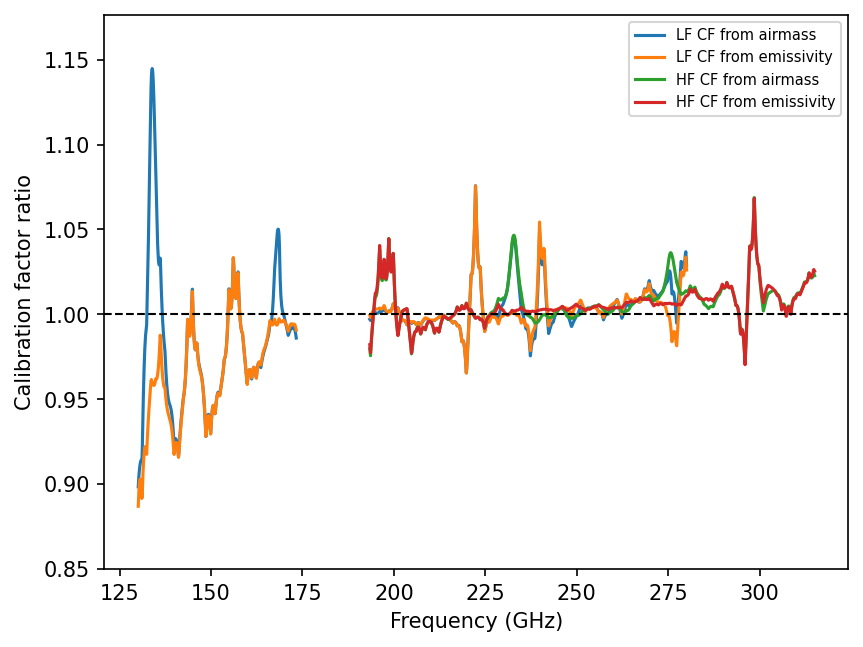}
\caption{
Recovery of the effective calibration spectrum $c'_\nu = k\,\bandpass$ from synthetic data. 
The figure shows the ratio between the calibration factors recovered with the airmass and emissivity methods and the input calibration spectrum used to generate the synthetic spectra, for the LF and HF bands. 
Perfect recovery corresponds to a ratio of unity. Both methods reproduce the input calibration spectrum to within a few percent across the usable frequency range. 
Deviations occur only near strong atmospheric absorption features (e.g. the 183\,GHz water vapour line and the 119\,GHz oxygen line), where the assumption of linear spectral variation with airmass or emissivity breaks down due to the convolution of sharp atmospheric structures with the instrumental sampling. 
These frequency intervals are excluded from the scientific analysis in the main text.
}
\label{fig:synth_cf}
\end{figure}

\subsection{Scope and limitations}

This appendix demonstrates that the emissivity and airmass calibration methods are mathematically consistent and unbiased under controlled conditions where the assumptions of the forward model are exactly satisfied.

The synthetic validation isolates the calibration formalism by excluding instrumental noise, bandpass uncertainties, stray-light variations, and elevation-dependent optical deformations. 
Consequently, any deviations observed in real data must arise from additional physical or instrumental effects rather than from intrinsic limitations of the calibration method itself.

In practice, residual departures from ideal behaviour may result from atmospheric variability within a scan, imperfections in the optical path difference reconstruction, or elevation-dependent mechanical effects. 
These effects are addressed and quantified in the main body of the paper.

The present validation therefore confirms that, provided the linear emissivity regime holds, the calibration framework can recover the effective calibration spectrum directly from science observations without requiring prior knowledge of the bandpass shape.

\end{document}